\documentclass[11pt,a4paper]{article}
\usepackage{booktabs,tabularx}
\usepackage[margin=2.5cm]{geometry}
\usepackage{amsmath,amssymb,amsthm}
\usepackage{bm}
\usepackage{graphicx}
\usepackage{microtype}
\usepackage{xcolor}
\usepackage{natbib}         
\usepackage{float}
\usepackage{hyperref}
\hypersetup{colorlinks=true,linkcolor=blue!45!black,citecolor=blue!45!black,urlcolor=blue!45!black} 
\newcommand{\GL}{\mathrm{GL}}
\newcommand{\OG}{\mathrm{O}}

\newcommand{\Syx}{\Sigma_{yx}}

\newcommand{\Fisher}{\mathcal{I}}

\newcommand{\rank}{\operatorname{rank}}

\newcommand{\todo}[1]{\textcolor{red}{[\textbf{TODO:} #1]}}

\newtheorem*{claim}{Logical argument}

\title{\bfseries Informative Fisher Modes in Neural Networks as a Result of Explicit Symmetry Breaking}

\author{%
I.~T.~Ersoy\thanks{Corresponding author:
\href{mailto:talha.ersoy@uni-potsdam.de}
{talha.ersoy@uni-potsdam.de}}
\quad B.~Ladewig
\quad K.~Wiesner\\[4pt]
\normalsize Institute of Physics and Astronomy,
University of Potsdam}
\date{\today}

\begin{document}
\maketitle

\begin{abstract}

\noindent Learning in a neural network takes place on a high dimensional, non-convex  loss landscape,  the Hessian curvature of which dictates the learning dynamics. Despite its high dimensionality, the large number of architectural symmetries explains some of the geometric  of this landscape, in particular flat directions, i.e. near-zero modes of the Hessian. These architectural symmetries have therefore been suggested as an organizing principle of deep learning. It remains an open question, though, wether there are additional organizing principles in the remaining degrees of freedom in the learning dynamic. Here we identify task-dependent contingent symmetries as such an additional organizing principle. We show that  non-flat learning directions emerge due to explicit breaking of these contingent symmetries. We isolate this symmetry in an analytically solvable minimal model. Furthermore, we show that, by tuning an L2 penalty, signatures of the symmetry breaking become measurable. As a consequence of the same mechanism, adding noise to the data drives the model through a sequence of broken symmetry recoveries in the form of Baik-Ben-Arous-Péché (BBP) transitions. We demonstrate analytically for linear and numerically for deep linear networks that the contingent symmetry group remains explicit, and that the symmetry-breaking mechanism identifies which Hessian modes carry task-specific information. These signatures are measurable without a symmetry-group description and, hence, offer a route to study symmetry breaking as a more general principle in nonlinear networks and other learning systems.
\end{abstract}

\tableofcontents

\section{Introduction}

\label{sec:intro}
Understanding how neural networks learn and generalize is a central problem in machine learning, statistics, and physics. Research approaches this problem through optimization, statistical inference, and the geometry of high-dimensional loss landscapes. Statistical physics provides complementary concepts, including symmetries, fluctuations, and phase transitions. A common challenge is to connect task structure with the behavior of the learning system. Here we examine how task constraints restrict model freedom and when the resulting curvature in the loss landscape represents statistical information. Deep linear networks trained from small initialization acquire dominant singular directions on different timescales \citep{saxe}. In nonlinear networks, generalization can improve abruptly after a long period of apparently successful fitting \citep{power2022grokking,grokking}. Physical learning systems also develop structured responses as they adapt: their task performance and mechanical or electrical response become linked through changes in the relevant curvature operators \citep{stern-supervised,stern-effects,stern-become}. A tractable model can isolate each feature constraint, the freedom it removes, and the resulting curvature. How to connect task structure with the behavior of the learning system is addressed by different communities using different methodologies. These include information and differential geometry, singular learning theory, statistical physics, theory of criticality and random matrix theory (RMT). Our work builds on all of these to some degree. We summarize the key findings, relevant for this work, in the following section. \\

\subsection{Prior work}
\emph{One line of work} investigated the geometry of DNN error landscapes by varying the strength of an $L_2$ penalty. \citet{ziyin-ueda} derived an \emph{onset of learning} as the $L_2$ penalty decreases: the transition is continuous with one hidden layer and discontinuous at greater depth. \citet{cascade} showed analytically that in deep linear networks the onset is only the first of a cascade of transitions, each associated with a singular direction of the data. In nonlinear networks, \citet{ersoy1} identified further transitions at lower $L_2$ strengths, related them to changes in error-landscape curvature, and associated each phase with distinguishable aspects of the task, which they demonstrated both for synthetic   and for real data \citep{ersoy2}. In the same framework, metastability and noise-driven escape from local minima have also been investigated and found to be a possible mechanism for delayed generalization in nonlinear DNNs \citep{grokking}. Taken together, these results reveal  a sequence of learning transitions, each associated with a geometric location and a Hessian spectrum signature.
\\
\emph{A second line of work} examined Hessian spectra of error landscapes at points of changing curvature directly. Empirical studies reported many small Hessian eigenvalues together with a smaller set of large eigenvalues \citep{sagun2018empirical}; random-matrix approaches explained aspects of the spectral structure and its dependence on the task and the parameter settings \citep{pennington2017geometry,bray2007statistics}. In spiked random-matrix models, the Baik-Ben-Arous-Péché (BBP) transition and related results described when signal-associated outliers separate from the bulk and when their eigenvectors correlate with the underlying signal \citep{baik2005phase,benaych2011eigenvalues}. Information geometry supplies a complementary local description through Fisher information \citep{amari,martens}. \citet{kuhn2026explaining} explain near-zero Hessian eigenvalues through the weak explicit breaking of architectural symmetries by nonlinearities and analyze the corresponding Fisher geometry.
\\
\noindent \emph{A third line of work} studies symmetries in neural-network parameter space and their consequences for learning, reviewed by \citet{zhao2025symmetry}. \citet{ziyin-sym} proved that mirror symmetries define fixed subspaces preserved by gradient descent and give the Hessian a corresponding block structure at symmetric configurations. Under additional assumptions, sufficiently strong weight decay favors minima in these subspaces; when the symmetry preserves the observation distribution, Fisher information is singular at the corresponding symmetric configurations. Symmetric configurations also reduce the expressivity of the linearized model and the dimension accessible to training \citep{ziyin-expressivity}. In a recent position paper, \citet{zxpc} proposed symmetry breaking and restoration as organizing principles for hierarchical learning.  Architectural symmetries hold independently of the data, whereas symmetries of a loss can depend on the task. In deep linear networks, architectural reparametrizations have a tractable geometry \citep{menon-geometry}, and an $L_2$ penalty selects balanced parametrizations of a fixed function in the full-rank square case \citep{lindsey-menon}. The contingent freedoms studied here preserve an auxiliary partial task but need not preserve the full objective.
\\
\subsection{Our Contribution}
We ask: \emph{How does imposing a task feature remove a task-preserving freedom and create a positive Fisher mode, and how does regularization affect this relation?} Answering this requires distinguishing a freedom left open by a partial task from a symmetry of a fixed objective and from a degeneracy at a particular parameter setting. It also requires separating the curvature of the data error from that of the penalized objective. Our contributions are fivefold. (i) We distinguish contingent symmetries, the freedoms a partially specified task leaves open, from architectural symmetries and task invariances, and introduce the feature filtration to isolate them. (ii) At a common regular exact solution, we show that imposing one independent feature constraint removes one such freedom and increases the Fisher rank by one, while architectural freedom survives. (iii) We show that an $L_2$ penalty can make reparametrization directions massive while they remain null directions of the Fisher metric: these penalty-lifted directions do not correspond to learning a feature. (iv) We combine the regularization cascade with spectral recovery thresholds in a learnability phase diagram over regularization strength and BBP noise amplitude, whose regions have the feature-subset labels of the filtration; for a uniform penalty the two probes order the features identically, while unequal penalty weights make the orderings compete and the
boundaries cross. (v) We state the observables by which the account can be tested in networks whose symmetry group is unknown. Section~\ref{sec:symmetry} defines the framework and filtration. Section~\ref{sec:breaking} develops the minimal example and its spectral interpretation. Section~\ref{sec:detecting} examines regularization and statistical recovery as probes of the resulting structure. Section~\ref{sec:where} extends the analysis to deep linear networks, and Section~\ref{sec:discussion} discusses nonlinear networks, other learning systems, and open questions.

\section{Conceptual framework and method}
\label{sec:symmetry}
The principle investigated here is that learning a feature restricts the choices left open by a task. To examine this idea, we need to distinguish the types of degrees of freedom and specify how a feature enters the analysis. We construct a sequence of partial tasks and compare their admissible solutions, symmetries, error curvature, and statistical distinguishability on the parameter space. This section defines the construction and the quantities to be compared. The relation between them is worked out in the minimal model of Sec.~\ref{sec:breaking}.

\subsection{Architectural, task, and contingent symmetries}
\label{sec:threetypes}
Let $\{f_\theta:\theta\in\Theta\subseteq\mathbb R^D\}$ be the model family. A configuration $\theta$ specifies the realized function $f_\theta$; the task specifies the requirements on that function. When a probability distribution is intended, we refer explicitly to the observation model $p_{\theta,k}$ introduced below. Consequently, two parameter settings can be equivalent for different reasons. We distinguish three sources of equivalence by what remains unchanged and by the conditions under which the equivalence holds.

\paragraph{Architectural symmetries.}
These are reparametrizations that leave the realized function unchanged:
\begin{equation}
  f_{g\cdot\theta}(x)=f_\theta(x)
  \quad\text{for every admissible }x.
  \label{eq:architectural-equivalence}
\end{equation}
They follow from the parametrization and hold independently of the task. For a two-layer linear network $f_\theta(x)=W_2W_1x$, the $w$ coordinates of the intermediate vector $W_1x$ are its hidden units and $w$ is its hidden width. With end-to-end map $M=W_2W_1$, the reparametrizations include
\begin{equation}
  (W_2,W_1)\longmapsto(W_2G,G^{-1}W_1),
  \qquad G\in\GL(w).
  \label{eq:internal}
\end{equation}
Hidden-unit permutations are another example; homogeneous activations also permit suitable rescalings. The available transformations depend on the architecture. The fibers of a factorized linear network and their group actions provide a rigorous setting for this redundancy \citep{menon-geometry,lindsey-menon}.

\paragraph{Data and task symmetries.}
These express distinctions that the fixed task does not require the model to resolve. For example, in a classification task an input transformation may preserve the assigned class label. Such an invariance is specified by the data distribution and the objective; it need not be a symmetry of an arbitrary model's predictions. To induce a symmetry of the learning objective, the corresponding transformation must preserve the loss and the relevant data measure. Its existence alone therefore does not establish a flat direction in parameter space. It remains part of the task when that task is fully specified.

\paragraph{Contingent symmetries.}
These preserve the features already imposed while allowing other aspects of the realized function to change. Their defining property is that a further feature constraint can distinguish configurations that were previously equivalent. They are the contingent symmetries investigated here. In linear networks some admit descriptions through transformations of both model and data coordinates. Their essential content is the preservation of a partial task. They can change the represented function and need not survive a further feature constraint. This distinction concerns the origin of the equivalence, rather than three necessarily disjoint groups. A transformation can preserve both the architecture's function and every partial task. We therefore identify the architectural freedom first and then ask what additional freedom the partial task leaves available.

\paragraph{Task descriptions as coarse-grainings.}
Let $\mathcal F=\{f_\theta:\theta\in\Theta\}$ be the set of realizable functions, and define the realization map $\mathcal R:\Theta\to\mathcal F$ by $\mathcal R(\theta)=f_\theta$. A partial task retains $k$ scalar feature values $c_i(\theta)$, each depending only on the realized function. Its feature map is therefore well defined by
\[ C_k(f_\theta)  =\bigl(c_1(\theta),\ldots,c_k(\theta)\bigr), \qquad \mathcal C_k=C_k(\mathcal F).
\]
For prescribed target values $\eta_i$, the scalar error is obtained through
\[ \ell_k(z)=\frac12\sum_{i=1}^k(z_i-\eta_i)^2, \qquad E_k=\ell_k\circ C_k\circ\mathcal R.
\]
These maps give descriptions containing no more detail than their inputs (Fig.~\ref{fig:coarse-graining}). Architectural transformations satisfy $\mathcal R(g\cdot\theta)=\mathcal R(\theta)$. Contingent transformations preserve $C_k(\mathcal R(g\cdot\theta))=C_k(\mathcal R(\theta))$ while potentially changing the realized function.
The maps perform the coarse-graining; the transformations preserve the descriptions retained by those maps. Their orbits lie within the corresponding fibers, where a fiber is the set of points having the same image. A fiber need not be a single symmetry orbit. The final scalar error discards further distinctions: equal error does not generally imply equal feature values. 
\begin{figure}[tbp]
\centering
\[
 \underbrace{\Theta}_{\text{parameters}}
 \xrightarrow{\mathcal R}
 \underbrace{\mathcal F}_{\text{functions}}
 \xrightarrow{C_k}
 \underbrace{\mathcal C_k}_{\text{feature values}}
 \xrightarrow{\ell_k}
 \underbrace{\mathbb R_{\geq0}}_{\text{error}}
\]
\caption{\textbf{Successive descriptions of a model configuration.}
The realization map $\mathcal R$ sends a parameter configuration to its input-output function. Architectural symmetries change the parameters while leaving this function unchanged. The map $C_k$ retains only the $k$ feature values specified by the partial task. Contingent symmetries can change the function while preserving these values; their parameter-space orbits lie within fibers of $C_k\circ\mathcal R$. The final map $\ell_k$ compares the retained values with their targets and summarizes the discrepancies by a scalar error. Equal scalar errors need not correspond to equal feature values. Adding a feature replaces $C_k$ by the finer description $C_{k+1}$, allowing some previously equivalent functions to be distinguished.}
\label{fig:coarse-graining}
\end{figure}

\subsection{Feature filtration as an analytical construction}
\label{sec:filtration}
In the tractable construction, the task consists of $r$ \emph{features}: components of the target that the model has to represent, with prescribed strengths $\eta_i$, $i=1,\ldots,r$. We use the word only in this sense. The model's coefficient along feature $i$ is a scalar functional $c_i(\theta)$; \emph{imposing} feature $i$ means requiring $c_i(\theta)=\eta_i$, and the model \emph{represents} feature $i$ when $c_i(\theta)\neq0$.
Each $c_i$ depends on the realized function, so architectural reparametrizations leave it unchanged. The conditions $c_i(\theta)=\eta_i$ express what it means to represent the corresponding features correctly. For an aligned linear model, the desired linear input-output map $S=\sum_i\eta_i u_i v_i^{\top}$ and the represented map $M=\sum_i p_i u_i v_i^{\top}$ use the same fixed orthonormal input and output directions. Then $c_i(\theta)=u_i^{\top}M(\theta)v_i=p_i$. This is what we mean by alignment; no such decomposition is assumed for every learning problem. Choose an ordering of the constraints and define the auxiliary errors
\begin{equation}
  E_k(\theta)=\frac12\sum_{i=1}^{k}
  \bigl(c_i(\theta)-\eta_i\bigr)^2,
  \qquad k=0,\ldots,r,
  \label{eq:filtration-loss}
\end{equation}
with $E_0=0$. The \emph{full task} imposes all $r$ constraints; a \emph{partial task} imposes only the specified subset, here the first $k$. An \emph{exact fit} of that partial task satisfies every retained equality, so $E_k=0$. Assume that the complete set is jointly realizable. Its partial-task solution sets, equivalently the global minimum sets of $E_k$, are then
\begin{equation}
  \mathcal M_k=
  \{\theta:c_i(\theta)=\eta_i\text{ for }i\leq k\},
  \qquad
  \Theta=\mathcal M_0\supseteq\mathcal M_1
  \supseteq\cdots\supseteq\mathcal M_r.
  \label{eq:filtration-minima}
\end{equation}

This nesting corresponds to a refinement of task descriptions. Let $\phi_k:\mathcal C_{k+1}\to\mathcal C_k$ discard the last feature value. Then 
\[ C_k=\phi_k\circ C_{k+1}, \qquad k=0,\ldots,r-1. \]
For $k=0$, the retained description is the single empty tuple. Functions with identical first $k+1$ feature values therefore have identical first $k$ values, whereas the converse need not hold. Discarding a feature coarse-grains the description; adding a feature refines it. The exact-solution sets are its target fibers,
\[ \mathcal M_k =(C_k\circ\mathcal R)^{-1} \bigl((\eta_1,\ldots,\eta_k)\bigr). \]
This nesting is part of the construction: each partial task retains the constraints already imposed and adds one further requirement. Once the coefficient functionals, their target values, and their ordering are fixed, the chain is determined. Different orderings can give different intermediate sets. These are constraint sets, not generally vector subspaces or nested sets of exactly different ranks. The remaining coefficient functionals are absent from $E_k$; their values may vary to the extent allowed by the architecture and the retained constraints. Whether these choices admit a group description, and what changes when the next constraint is imposed, are questions for the subsequent analysis. The level $k=0$ means that no feature constraint has yet been imposed. It does not mean that every parameter setting realizes the zero function. Similarly, an unconstrained feature need not have a zero coefficient. This distinction separates the auxiliary partial task from a model that has failed to fit part of a fixed, fully specified task. The filtration is an analytical device for isolating the effect of a feature constraint. Its ordering does not prescribe the arrival of data or the trajectory of an optimizer. Whether regularization or statistical recovery produces a corresponding sequence is examined separately in the later sections.

\subsection{Symmetry, curvature, and information}
\label{sec:group}
\label{sec:filtration-information}
For each partial task we examine which transformations preserve its features, how the error responds to parameter perturbations, and which perturbations change the associated observation distribution. These are distinct descriptions, whose relationship must be established for the model under consideration. To specify the symmetry question, fix a common group $\mathcal D$ of admissible invertible transformations of parameter space and define
\begin{equation}
  G_k=\{g\in\mathcal D:
  c_i(g\cdot\theta)=c_i(\theta)
  \text{ for all }\theta\text{ and }i\leq k\}.
  \label{eq:filtration-group}
\end{equation}
The orbit $G_k\cdot\theta$ is the set of configurations reached by these transformations from $\theta$. When the action is smooth, a one-parameter subgroup $g_t$ has generator $X(\theta)=\left.\frac{d}{dt}(g_t\cdot\theta)\right|_{t=0}$. The analysis will determine the groups and their orbits, and compare the independent tangent directions generated at a solution. We do not assume that an orbit is the entire minimum set or that every generator produces a nonzero tangent direction. We use \emph{explicit symmetry breaking} for a change in the invariance of the objective and \emph{spontaneous symmetry breaking} for the selection of a configuration with less symmetry than an unchanged objective \citep{goldenfeld,altland}. In the filtration, adding a feature constraint explicitly reduces the contingent group, $G_k\to G_{k+1}$, because the objective changes from $E_k$ to $E_{k+1}$. A one-parameter action that leaves a configuration $\theta$ unchanged has $X(\theta)=0$ and supplies no nonzero orbit tangent there. The distinct symmetry effects of regularization are introduced with the penalized objective in Sec.~\ref{sec:l2}. 
For the curvature calculation, define
\begin{equation}
  H_k(\theta)=\nabla_\theta^2E_k(\theta).
  \label{eq:framework-hessian}
\end{equation}
At a stationary point, $v^{\top}H_kv$ for a unit parameter variation $v$ is the directional curvature of the error $E_k$. We call its zero-eigenvalue modes \emph{flat} and its positive-eigenvalue modes \emph{massive}. In the regularized sections we distinguish $H_E=\nabla^2E$ from $H_{\mathcal L}=\nabla^2\mathcal L_\beta$. Fisher information is a separate quadratic form. A Hessian zero mode need not arise from an exact invariance. We also use the coefficient Jacobian $J_k=\partial(c_1,\ldots,c_k)/\partial\theta$ to describe which coefficient values change under an infinitesimal parameter variation. The relation between this Jacobian, symmetry generators, and the Hessian is derived in Sec.~\ref{sec:breaking}. To specify the information calculation, associate the partial task with independent Gaussian observations
\begin{equation}
  z_i=c_i(\theta)+\varepsilon_i,
  \qquad
  \varepsilon_i\sim\mathcal N(0,\varsigma^2),
  \qquad i=1,\ldots,k,
  \label{eq:framework-observations}
\end{equation}
with fixed variance. Let $p_{\theta,k}(z)$ denote their joint density. The Fisher information is
\begin{equation}
  \Fisher_k(\theta)=
  \mathbb E_{z\sim p_{\theta,k}}\!\left[
    \nabla_\theta\log p_{\theta,k}(z)
    \nabla_\theta\log p_{\theta,k}(z)^{\top}
  \right].
  \label{eq:framework-fisher}
\end{equation}
It measures the local statistical distinguishability of nearby parameter settings within the specified observation model \citep{bishop,amari,martens}. We call the pullback quadratic form $v^{\top}\Fisher_kv$ the \emph{Fisher metric}; it may be degenerate in redundant parameter coordinates. We compare its null and positive directions with those of the error Hessian. Their agreement is to be established under the stated conditions; it is not part of the definition. The distinction matters once a penalty is added: its curvature can make a direction massive without introducing additional statistical distinguishability.

\subsection{Assumptions and analysis strategy}
\label{sec:setup}
The exact local analysis assumes jointly realizable constraints, twice continuously differentiable coefficient functionals, and \emph{regular exact solutions}: points of $\mathcal M_k$ at which $\rank J_k=k$, so the retained constraint differentials are independent. Smooth group actions are used where symmetry generators are calculated. The Gaussian observation model fixes the statistical meaning of the Fisher calculation. Singular configurations, dependent constraints, and settings without an explicit group description require separate treatment. Section~\ref{sec:breaking} applies this construction to a two-feature factorized model and computes its minimum sets, symmetry actions, and curvature and information modes. We then keep the task fixed and examine how an $L_2$ penalty competes with its feature requirements (Sec.~\ref{sec:l2}) and which curvature it adds without information(Secs.~\ref{sec:l2} and~\ref{sec:sectors}). The BBP connection examines how finite noisy data resolve feature directions to different degrees, providing a statistical counterpart to exact feature inclusion (Sec.~\ref{sec:bbp}), and the two probes combine into a learnability phase diagram (Sec.~\ref{sec:phase}). The later extensions distinguish the results established for linear networks from the proposed implications for nonlinear networks and other learning systems.

\section{Explicit symmetry breaking produces positive Fisher information}
\label{sec:breaking}

We now realize the construction of Sec.~\ref{sec:symmetry} in a model with four parameters and two features. Its purpose is to make the same event visible in three descriptions: each added feature constraint reduces the minimum set, removes a task-preserving freedom, and creates a positive Hessian and Fisher mode. The architectural freedom to represent a fixed function remains throughout. For the derivation of the explicit symmetry breaking when features are introduced and the resulting flat-to-massive transition in the spectra of the error Hessian and the Fisher metric, we use the filtration scheme introduced in Sec.~\ref{sec:filtration}.

\subsection{A minimal model and the partial tasks}
\label{sec:minimal}

\begin{figure}[H]
  \centering
  \includegraphics[width=\textwidth]{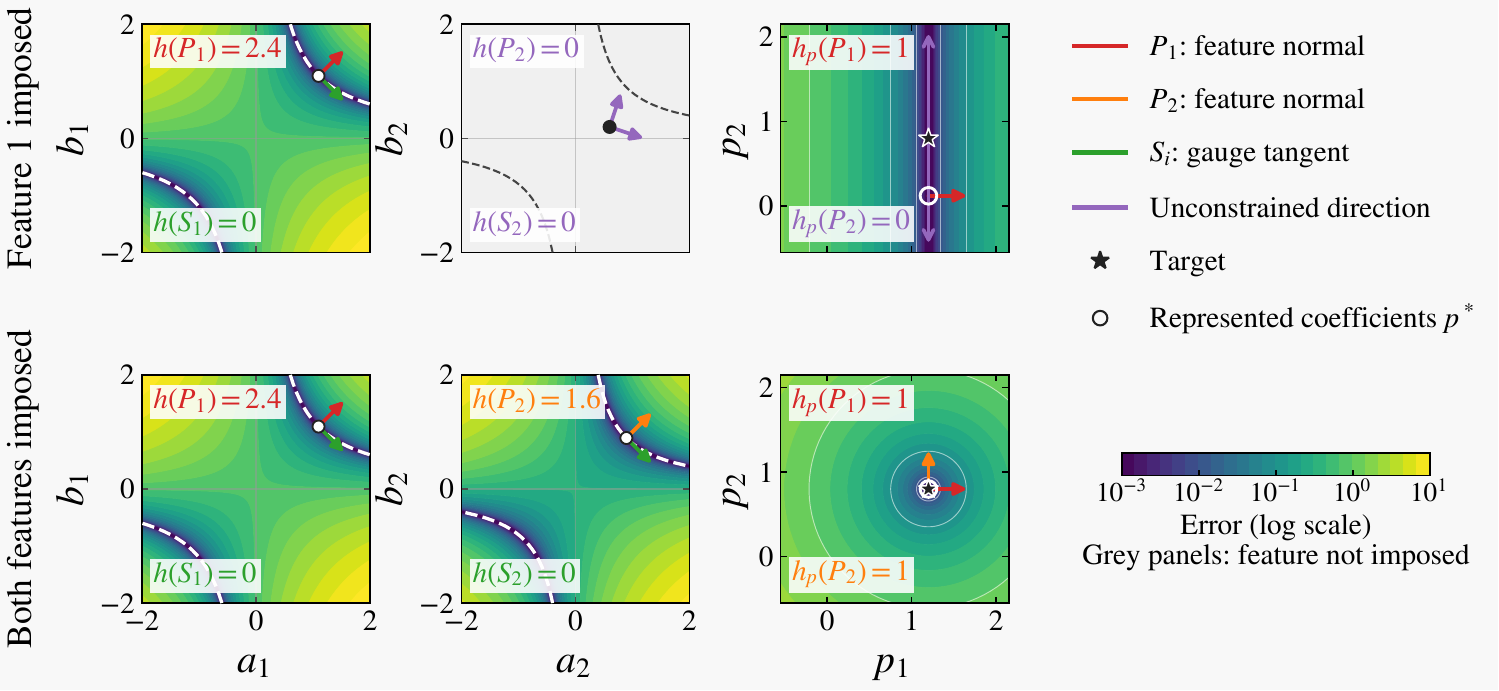}
  \caption[Imposing features one at a time]{%
    \textbf{Imposing features one at a time.}
    Each row shows the loss $L=\sum_{i=1}^{2}\tfrac12\rho_i(a_ib_i-\eta_i)^2$ with targets $\eta=(1.2,\,0.8)$,
    in the factor coordinates $(a_1,b_1)$ (left) and $(a_2,b_2)$ (centre),
    and in the coefficient coordinates $p_i=a_ib_i$ (right).
    \emph{Top row:} only feature~1 is imposed, $(\rho_1,\rho_2)=(1,0)$.
    \emph{Bottom row:} both features are imposed, $(\rho_1,\rho_2)=(1,1)$.
    Colours show the error on a shared logarithmic scale.
    A grey panel marks a feature that is not imposed, so the loss does not depend on its factors.
    \par
    \emph{Factor panels:} for an imposed feature, the dashed white hyperbola $a_ib_i=\eta_i$ is the set of exact minima,
    and the white dot is the balanced representative $a_i=b_i=\sqrt{\eta_i}$.
    The arrow $P_i$ is normal to the hyperbola and changes $p_i$;
    its Hessian eigenvalue is $h(P_i)=\rho_i(a_i^2+b_i^2)=2\rho_i\eta_i$.
    The arrow $S_i$ is tangent to the gauge orbit $(a_i,b_i)\mapsto(g\,a_i,\,b_i/g)$, which leaves $p_i$ unchanged, so $h(S_i)=0$.
    For a feature that is not imposed, both directions are unconstrained (purple),
    the dashed grey hyperbola is shown only for reference,
    and the black dot marks an arbitrary configuration.
    \par
    \emph{Coefficient panels:} the star marks the target $\eta$ and the open circle the represented coefficients $p^*$.
    An imposed coefficient has curvature $h_p(P_i)=\rho_i$.
    A coefficient that is not imposed is a flat direction (purple double arrow), so $p_2$ is undetermined in the top row.
    With both features imposed, $p^*$ coincides with the target.}
  \label{fig:filtration}
\end{figure}

The minimal model has four parameters $\theta=(a_1,b_1,a_2,b_2)$ and two coefficients, one per feature. Its realized predictor is $f_\theta(x)=p_1x_1+p_2x_2$, with
\begin{equation}
  c_i(\theta)=p_i=a_i b_i,\qquad i=1,2,
  \label{eq:minimal}
\end{equation}
Take targets $\eta_1>\eta_2>0$. Thus $E_1=\tfrac12(p_1-\eta_1)^2$ and $E_2=\tfrac12\sum_{i=1}^2(p_i-\eta_i)^2$. For whitened inputs, $\mathbb E[xx^{\top}]=I$, the latter is also the population error against the noiseless target $f_*(x)=\eta_1x_1+\eta_2x_2$: $E_2=\tfrac12\mathbb E_x[(f_\theta(x)-f_*(x))^2]$. Here \emph{population error} means an expectation over the input distribution, rather than an average over a finite sample. The diagonal factorization excludes hidden-unit mixing; the full matrix model is introduced in Sec.~\ref{sec:where}. The architectural symmetry \eqref{eq:internal} reduces here to
\begin{equation}
  (a_i,b_i)\;\longmapsto\;(g\,a_i,\ b_i/g), \qquad g\in\GL(1),
  \label{eq:minimal-gauge}
\end{equation}
under which $p_i$ is unchanged. For nonzero products its orbits are the hyperbolas $a_ib_i=\mathrm{const}$ in the $(a_i,b_i)$ plane, all points on a given hyperbola have the same coefficient $p_i$: $E_k$ is constant along the curve and the Fisher metric has a null direction tangent to it. At an exact fit of an imposed feature, the transverse direction changes $p_i$ and has positive error curvature. Writing $\nabla p_i=(b_i,a_i)$ makes the pair explicit,
\begin{equation}
  \hat n \;=\; \frac{(b_i,\;a_i)}{\sqrt{a_i^2+b_i^2}}
  \quad\text{(changes $p_i$)},
  \qquad
  \hat k \;=\; \frac{(a_i,\,-b_i)}{\sqrt{a_i^2+b_i^2}}
  \quad\text{(preserves $p_i$)},
  \label{eq:minimal-dirs}
\end{equation}
and $\hat n\cdot\hat k=0$,  so the gauge tangent $\hat k$ and the normal $\hat n$ are orthogonal. They are respectively Hessian-flat and massive on the nonzero exact-fit hyperbola of an imposed feature; loss invariance along a curved orbit alone does not imply a zero ordinary Hessian eigenvalue away from stationarity. The two directions distinguish reparametrization along a fixed-product orbit from a change of the coefficient $p_i$. The contingent symmetry appears while a feature is unconstrained. Before the second constraint is imposed, nothing in the objective depends on $p_2$: at an exact fit of feature~1, every value of $p_2$ is compatible with the partial task, so varying it leaves $E_1=0$. Once $p_2=\eta_2$ is imposed, it does. Explicitly, let $\mathcal D$ consist of the independent rescalings $(a_i,b_i)\mapsto(g_ia_i,\,h_ib_i)$, $g_i,h_i\neq0$. The groups \eqref{eq:filtration-group} are then
\begin{equation}
  G_k=\{(g_1,h_1,g_2,h_2):\ g_ih_i=1\ \text{for } i\le k\},
  \qquad \dim G_k=4-k,
  \label{eq:minimal-groups}
\end{equation}
so $G_0\supset G_1\supset G_2$, with $G_2$ the architectural symmetry \eqref{eq:minimal-gauge} of both features. Through a point of $\mathcal M_k$ with all $a_i,b_i\neq0$, the $G_k$-orbit has dimension $4-k$, that of $\mathcal M_k$: it fills the solution set locally, and its tangents are exactly the null directions of the Hessian $H_k=\nabla^2E_k$ at that exact solution. The factor lost in the step $G_1\to G_2$, the rescaling of $p_2$ by $g_2h_2$, is the contingent symmetry that imposing the second feature breaks. The transition from one state to the other is the elementary breaking, and Fig.~\ref{fig:filtration} shows both sides of it.

\subsection{Reducing contingent symmetries generates a nested learning hierarchy}


\label{sec:chain}

The contingent groups form a nested chain
\begin{equation}
    G_0 \;\supseteq\; G_1 \;\supseteq\;\dots\;\supseteq\; G_{r},
\label{eq:chain}
\end{equation}
maximal when no feature is imposed ($k=0$) and minimal when all are ($k=r$). Imposing the $(k{+}1)$-th feature is the step $G_k\to G_{k+1}$:
the group loses transformations that change the new feature, and the set of realizations of the current accuracy, made up of orbits of equivalent parameter configurations, contracts to the set satisfying the additional constraint. Across the filtration this is explicit symmetry breaking in the sense of Sec.~\ref{sec:group}: the objective itself loses invariance. Note that the modes are independent, so the ordering need not be unique. Instead of considering the models that capture feature 1 and feature 2 as a subset of the models that only capture feature 1, we can construct the nesting, starting with feature 2.

\subsection{One independent constraint increases Hessian and Fisher rank by one}
\label{sec:rankone}
The added constraint detects a variation that previously left all retained coefficients unchanged. To count the resulting directions, consider a small parameter displacement $v$: the coefficient Jacobian $J_k$ maps it to the first-order changes of the first $k$ coefficients. At a regular exact solution, the displacements that remain tangent to the solution set are therefore precisely those annihilated by $J_k$,
\begin{equation}
  T_\theta\mathcal M_k=\ker J_k(\theta).
  \label{eq:filtration-tangent}
\end{equation}
If the first $k$ constraints have independent differentials, the local dimension of this set is $D-k$, and a further independent scalar constraint removes one tangent direction. At an exact solution of the first $k$ constraints,
\begin{equation}
  H_k=J_k^{\top}J_k,
  \qquad \ker H_k=\ker J_k.
  \label{eq:filtration-hessian}
\end{equation}
To compare successive levels at the same point, take $\theta\in\mathcal M_{k+1}$ and set $u=\nabla c_{k+1}(\theta)$. Then
\begin{equation}
  H_{k+1}=H_k+uu^{\top},
  \qquad
  v^{\top}H_{k+1}v=(u^{\top}v)^2
  \quad\text{for }v\in\ker H_k.
  \label{eq:filtration-curvature}
\end{equation}
An old flat direction acquires curvature precisely when it changes the new feature to first order. If $u$ is independent of the preceding constraint gradients, the number of positive Hessian eigenvalues increases by one and the remaining tangent directions stay flat; the positive eigenvectors can rotate unless the constraint gradients are orthogonal. A continuous symmetry that preserves the imposed features has a generator with $J_kX=0$, hence $H_kX=0$ at a fitted solution; after the additional constraint the same direction has curvature $(u^{\top}X)^2$. For the Gaussian observations
\eqref{eq:framework-observations} the expected Fisher information is
\begin{equation}
  \Fisher_k=\frac{1}{\varsigma^2}J_k^{\top}J_k,
  \qquad
  H_k=\varsigma^2\Fisher_k
  \quad\text{at exact fit},
  \label{eq:filtration-fisher}
\end{equation}
so the curvature gained in \eqref{eq:filtration-curvature} is a new direction of statistical distinguishability. In the minimal model the nonzero entries of the Jacobian row for feature $i$ is $\nabla p_i=(b_i,a_i)$, so imposing it adds exactly the massive direction $\hat n$ of \eqref{eq:minimal-dirs} and leaves $\hat k$ flat. Away from exact fit the error Hessian also contains residual-dependent terms, and a penalty contributes curvature of its own; neither can be identified with Fisher information (Sec.~\ref{sec:l2}). 

\subsection{Feature-induced massive modes are emergent learning directions}
\label{sec:emergent}

\noindent The result of Sec.~\ref{sec:rankone} identifies the same newly constrained parameter variation in the error Hessian and in the observation distribution. For a Gaussian model with fixed noise scale, the mean-squared error is the negative log-likelihood up to an additive constant and a positive factor \citep{bishop}: minimizing the loss is maximum-likelihood estimation. Consequently, the loss Hessian is proportional to the Hessian of the negative log-likelihood, whose expectation under the model is the Fisher-information matrix \citep{amari,martens}; the negative log-likelihood Hessian for the observed sample is the observed information, including when evaluated at an optimum \citep{observed}. For the feature observations of Eq.~\eqref{eq:framework-observations}, exact fit removes the residual term. For a correctly specified Gaussian population loss, fitting the conditional mean instead makes its expectation vanish. With matching sum or average conventions,
\begin{equation}
  \nabla^2 E\big|_{\text{exact fit}} \;=\; \varsigma^2\,\Fisher(\theta),
  \qquad
  \Fisher(\theta) = \mathbb{E}\!\left[\nabla\log p_\theta \,\nabla\log p_\theta^{\top}\right],
  \label{eq:hessian-fisher}
\end{equation}
where $\varsigma^2$ is the noise variance. On locally identifiable statistical coordinates, Fisher information defines a Riemannian metric; its pullback to redundant parameters is generally singular \citep{amari}: for a fixed parametrization, its large-eigenvalue directions are the parameter combinations most tightly constrained locally by the data, and its null directions have zero first-order statistical distinguishability. At singular points they may still be distinguishable at higher order. This turns the flat and massive directions of the filtration into statements about information. A symmetry direction is one along which the model distribution does not change, so by \eqref{eq:hessian-fisher} it is a null direction of $\Fisher$: the corresponding parameter combination is non-identifiable and carries zero information. Unbroken contingent and reparametrization symmetries give degenerate directions of the Fisher metric,  the setting of singular learning theory \citep{watanabe}. In the minimal model at exact solutions with all factors nonzero they give all of them, Eq.~\eqref{eq:minimal-groups}; in deep linear networks they need not (Sec.~\ref{sec:sectors}). Conversely, a direction that becomes massive when a feature is imposed, Eq.~\eqref{eq:filtration-curvature}, is a null direction of $\Fisher$ turning positive: that parameter variation now changes the observation distribution to first order, so its Fisher quadratic form is positive. This is local statistical distinguishability along a direction, not unique identification of the entire parameter vector. For an independent Gaussian channel $y_i=p_i x_i+\varepsilon_i$, with $x_i\sim\mathcal N(0,1)$ and noise variance $\varsigma^2$, the mutual information is $\tfrac12\log(1+p_i^2/\varsigma^2)$. This is a different quantity from parameter Fisher information. It vanishes when the represented coefficient is zero, not merely when the corresponding constraint is absent. For a fixed teacher channel with coefficient $\eta_i$, the input-output mutual information is fixed by the data distribution and does not turn on during training. We call the newly Fisher-positive direction in the auxiliary filtration an \emph{emergent learning direction}: an independent constraint makes a previously task-flat variation locally distinguishable. Penalty-lifted architectural directions remain Fisher-null and therefore do not become learning directions in this sense (Sec.~\ref{sec:l2}).

\section{Detecting learning directions}

\label{sec:detecting}
So far the symmetry has been identified and the consequences of its breaking computed, but nothing has been said about \emph{when} a breaking occurs. The chain \eqref{eq:chain} is a sequence of subgroups; it does not say what moves a system along it, nor whether every step in it is reachable. Two established phenomena answer these two questions, and in doing so give the principle empirical content. The first is the cascade of transitions driven by an $L_2$ penalty, which supplies a control parameter: it lets us compare which features are represented across penalized objectives and locate transitions along an ideal path of minimizers. The second is the detachment of a signal direction from a noise bulk, which supplies a precondition:
in the specified spiked ensemble, spectral recovery of a population feature requires its signal to exceed the relevant threshold. This is a condition for that recovery mechanism, not a universal impossibility result for every learner. Within the stated model, they compare feature activation with statistical resolution. Because the penalty also contributes curvature of its own, we separate the modes it introduces from the learning directions (Sec.~\ref{sec:l2}), and we close with a phase diagram in regularization strength and data-noise amplitude (Sec.~\ref{sec:phase}) as a temperature analog. That both are measurable without reference to the group is what makes the principle testable outside the setting where it was derived, and we return to that in Sec.~\ref{sec:neither}.

\subsection{The \texorpdfstring{$L_2$}{L2} cascade: regularization helps locate learning directions}

\label{sec:l2}

\begin{figure}[H]
  \centering
  \includegraphics[width=\textwidth]{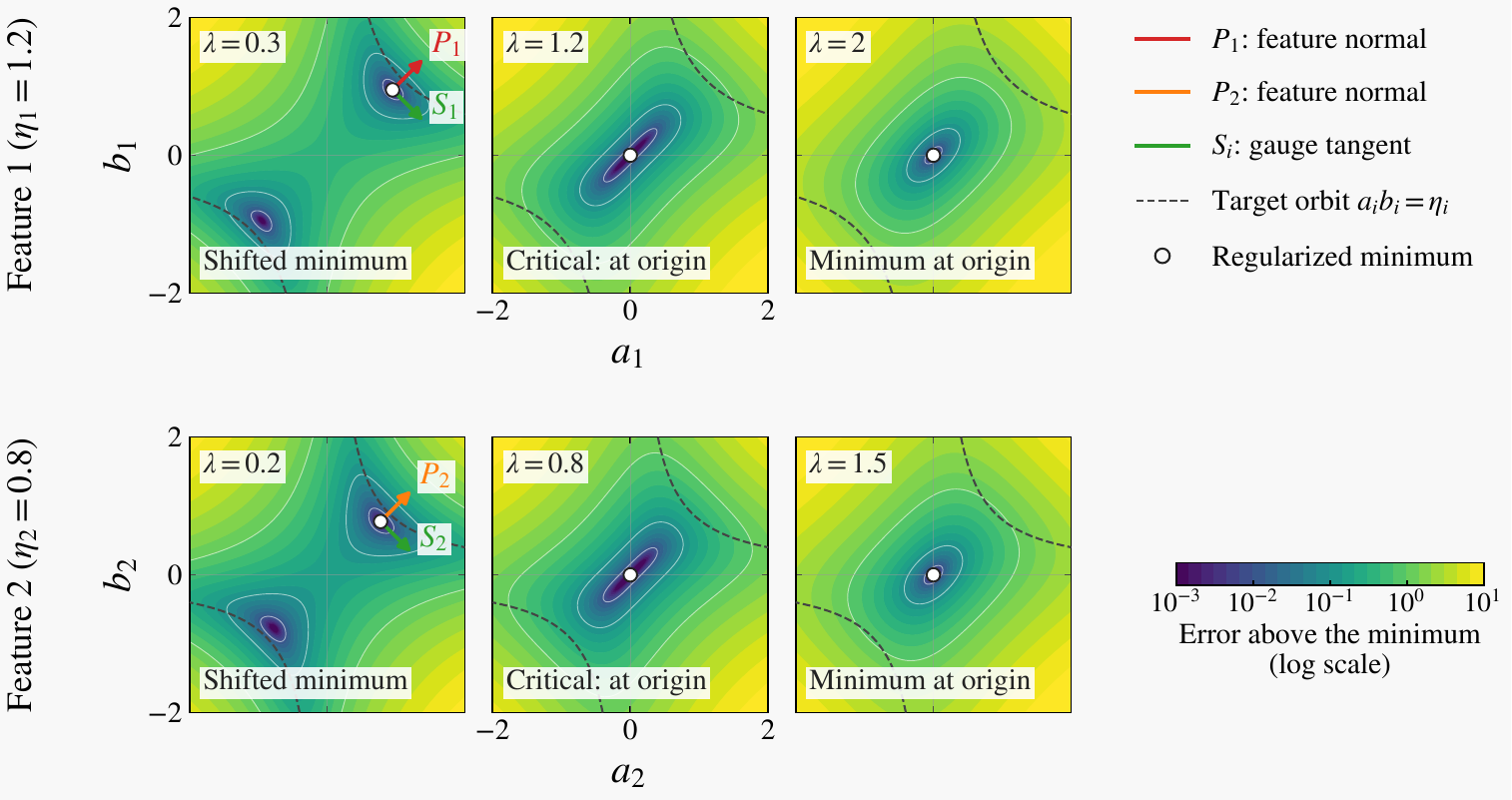}
  \caption[Regularization shrinks and ultimately suppresses each feature]{%
  \textbf{Regularization shrinks and ultimately suppresses each represented feature.} Each panel shows $\mathcal L_{\beta,i}=\tfrac12(a_ib_i-\eta_i)^2+\tfrac\beta2(a_i^2+b_i^2)$ in the factor plane $(a_i,b_i)$, with all feature weights $\rho_i=1$. The symbol $\lambda$ in the panels denotes the regularization strength $\beta$ used in the text. The top row shows feature~1 ($\eta_1=1.2$) and the bottom row feature~2 ($\eta_2=0.8$). Regularization increases from left to right, with $\beta=0.3,\,1.2,\,2$ in the top row and $\beta=0.2,\,0.8,\,1.5$ in the bottom row, as labelled in each panel. Colours show the loss above the panel's minimum on a logarithmic scale shared by all panels, so dark regions lie close to a minimum.
  \par
  The black dashed hyperbola $a_ib_i=\eta_i$ is the unregularized exact-fit set. For $0<\beta<\eta_i$ (left), the white dot marks the positive regularized minimum $a_i=b_i=\sqrt{\eta_i-\beta}$; the sign-related minimum $a_i=b_i=-\sqrt{\eta_i-\beta}$ appears as the second dark well in the opposite quadrant. At the marked minimum, the normal $P_i$ (red or orange) has $H_{\mathcal L}$ eigenvalue $2(\eta_i-\beta)$ and positive Fisher norm.
  The gauge tangent $S_i$ (green) has $H_{\mathcal L}$ eigenvalue $2\beta$ but remains Fisher-null.
  \par
  At $\beta=\eta_i$ (middle), the two minima merge at the origin and one Hessian eigenvalue vanishes. For $\beta>\eta_i$ (right), the origin is the unique minimum and the feature coefficient is zero. Reading from right to left therefore illustrates continuous feature activation in this two-factor model.}
  \label{fig:cascade}
\end{figure}

The chain \eqref{eq:chain} is a sequence of subgroups but says nothing about when a system moves along it. A control parameter supplies that. We add an $L_2$ penalty of strength $\beta$ to the error of the full task, $E=E_r$,
\begin{equation}
  \mathcal{L}_\beta(\theta) \;=\; E(\theta) + \frac{\beta}{2}\,\lVert\theta\rVert^2,\qquad \beta\ge0,
  \label{eq:regloss}
\end{equation}
so that in the minimal model an active minimum for $0<\beta<\eta_i$ sits at $a_i=b_i=\pm\sqrt{\eta_i-\beta}$ and is lost at the critical strength $\beta_c^{(i)}=\eta_i$. A regularization sweep varies $\beta$ while keeping the data error $E$ fixed. Analytically, we compare global minimizers $\theta_\beta\in\operatorname*{arg\,min}_\theta\mathcal L_\beta$. In a training experiment, the optimizer, initialization, sweep rate, and relaxation time determine which minima are reached. The resulting transitions have been characterized. They are observed in general, nonlinear networks \citep{ersoy1,ersoy2}, and noise-driven escape from metastable phases has been proposed as a mechanism for grokking in the models studied in \citet{grokking}. In the aligned deep-linear setting, the transitions are continuous for one hidden layer and first order for two or more \citep{cascade,ziyin-ueda}; depth alone does not establish this classification for arbitrary nonlinear architectures. In the aligned separable setting with sufficient width and distinct strengths, lowering $\beta$ activates features in descending order of $\eta_i$; increasing $\beta$ suppresses them in the reverse order (Appendix~\ref{app:thresholds}). Along this path the data error is fixed, while the penalized objective changes and is not invariant under the full contingent group, so it does not implement the explicit contingent symmetry reduction of the filtration; it selects which features are represented. Its phase with features $1,\dots,k$ learned corresponds to level $k$ of the filtration: the same features, with active coefficients $p_i=\eta_i-\beta$ in this two-factor model. For each factor pair the penalty reduces the reciprocal-scaling symmetry $\GL(1)$ to $\OG(1)\simeq\mathbb Z_2$. This surviving sign symmetry, $(a_i,b_i)\mapsto(-a_i,-b_i)$, is architectural because it preserves $p_i$; it is not the contingent freedom to change an unconstrained coefficient. A nonzero minimum spontaneously selects one of its two sign-related representatives: above $\beta_c^{(i)}$ the minimum $a_i=b_i=0$ is invariant under it, below it the two minima $a_i=b_i=\pm\sqrt{\eta_i-\beta}$ are exchanged by it. Each transition can therefore be labelled by the same added feature as a step $G_k\to G_{k+1}$ of the auxiliary filtration (Sec.~\ref{sec:chain}); it is not itself that explicit change of task symmetry.
This is the first thing the cascade establishes about the principle: 
the feature ordering has a corresponding sequence of transitions in this solvable model, and the locations are computable in the solvable case and measurable in general. The second is that, for a uniform penalty, the ordering is forced rather than incidental. Because the $\eta_i$ then order the $\beta_c^{(i)}$, the levels are reached in a definite sequence, and the continuous sweep of global minimizers cannot skip a feature threshold in this separable model, which is the content of Sec.~\ref{sec:nested}, seen from the side of the control parameter. We use \emph{quasi-static global-minimum path} for this ideal comparison of global minimizers at successive $\beta$. It need not be reproduced by slowly continuing one training run: following a local minimum can retain a metastable state and produce hysteresis. At finite Langevin temperature a trajectory can also visit both sign-related minima; the coefficient $p_i$ is invariant under this sign exchange. The penalty also lifts directions that remain Fisher-null, and the minimal model shows both kinds of mass side by side. At an active balanced minimum the reparametrization direction $\hat k$ of Eq.~\eqref{eq:minimal-dirs} 
acquires curvature $2\beta$ in $H_{\mathcal L}$ (Fig.~\ref{fig:cascade}), yet $\nabla p_i\cdot\hat k=0$, so it remains a null direction of the Fisher metric: a pseudo-Goldstone mode of the $\GL(1)$ symmetry that the penalty breaks explicitly.
Before the feature is learned, at $a_i=b_i=0$, the Hessian of $\mathcal L_\beta$ has eigenvalues $\beta\pm\eta_i$ while the Fisher metric vanishes; the smaller one goes soft exactly at $\beta_c^{(i)}=\eta_i$, and below it the direction $\hat n$ becomes Fisher-positive. The penalty-lifted gauge mode is thus Fisher-null, whereas the normal $\hat n$ of an active feature is Fisher-positive. This does not identify all penalty curvature with Fisher-null directions. Continuous hidden-unit rotations in the full matrix model add a symmetry-generated flat sector; the full classification is in Sec.~\ref{sec:sectors}.
{Table~\ref{tab:model-scope} keeps the unregularized toy result, its penalized counterpart, and the full matrix calculation separate.

\begin{table}[tbp]
\centering
\small
\setlength{\tabcolsep}{5pt}
\renewcommand{\arraystretch}{1.18}
\caption{The three settings and the curvature statements established in each.}
\label{tab:model-scope}
\begin{tabularx}{\textwidth}{@{}>{\raggedright\arraybackslash}p{0.22\textwidth}>{\raggedright\arraybackslash}X>{\raggedright\arraybackslash}X@{}}
\toprule
Setting & Symmetry or control & Hessian and Fisher statement \\
\midrule
Bilinear partial tasks $E_k$ (Sec.~\ref{sec:breaking}) &
An added constraint reduces contingent freedom; reciprocal rescalings survive. &
At a common regular exact fit, $H_k=\varsigma^2\Fisher_k$; each independent scalar constraint increases both ranks by one. \\
\addlinespace
Bilinear $\mathcal L_\beta$ (Sec.~\ref{sec:l2}) &
The full task is fixed. Positive $\beta$ leaves the architectural sign symmetry $\mathbb Z_2$ per feature. &
At an active minimum, the gauge tangent has $H_{\mathcal L}$ eigenvalue $2\beta$ and zero Fisher norm. At the origin, onset eigenvalues are $\beta\pm\eta_i$ while Fisher information is zero. \\
\addlinespace
Full two-factor matrix $\mathcal L_\beta$ (Sec.~\ref{sec:sectors}) &
The architectural group retained by the norm penalty is $\OG(w)$. Hidden-unit mixing is allowed. &
Nonzero orbit tangents give flat modes of $H_{\mathcal L}$. The four-sector formulas apply under the stated one-hidden-layer assumptions. \\
\bottomrule
\end{tabularx}
\end{table}
}

\subsection{Noise tuning: BBP transitions give a smooth filtration}

\label{sec:bbp}
The second detector asks a different question. A spectral estimator can recover a population feature only when its direction remains resolvable against the fluctuations of the empirical cross-covariance. The threshold below concerns this estimator in a specified large-matrix ensemble. Random matrix theory makes this sharp. 
For a linear network trained with squared loss on empirically whitened inputs, the parameter-dependent part of the loss depends on the targets through the empirical input--output cross-covariance $\hat\Sigma_{yx}=\frac1N\sum_{\mu=1}^{N}y_\mu x_\mu^{\top}$, an $n_{\mathrm{out}}\times n_{\mathrm{in}}$ matrix estimated from $N$ samples. For an empirically whitened fixed design $X$, with $XX^{\top}/N=I$, and labels $Y=SX+\varepsilon$ with independent Gaussian noise entries of variance $\sigma^2$, the conditional cross-covariance is a spiked matrix,
\begin{equation}
  \hat\Sigma_{yx} = S + T Z, \qquad S=\sum_i \eta_i\,u_i v_i^{\top}, \qquad T=\sigma\sqrt{n_{\mathrm{in}}/N}.
  \label{eq:spiked}
\end{equation}
Here the target map $S$ specifies the conditional mean $\mathbb E[y\mid x]=Sx$; it equals the population cross-covariance when $\mathbb E[xx^{\top}]=I$. Without empirical whitening, the signal term is $S\hat\Sigma_{xx}$ and the conditional noise covariance also depends on $\hat\Sigma_{xx}$. The signal map $S$ has fixed low rank, its nonzero singular values are distinct feature strengths $\eta_i$, and its singular vectors specify the feature directions.

The noise $TZ$ is the fluctuation that finitely many noisy samples add: $Z$ has independent zero-mean entries of variance $1/n_{\mathrm{in}}$, and the amplitude $T$ grows with the label noise and falls with the number of samples. The results below hold for large dimensions at fixed aspect ratio $q=n_{\mathrm{out}}/n_{\mathrm{in}}\le1$; we restrict the formulas to this variance and aspect-ratio convention. Without a signal, the singular values of $\hat\Sigma_{yx}$ fill a bulk with upper edge $T(1+\sqrt q)$. A spike of strength $\eta$ separates from it only when its signal-to-noise ratio $\gamma=\eta/T$ exceeds
\begin{equation}
  \gamma_c = q^{1/4},
  \label{eq:bbp-threshold}
\end{equation}
the BBP transition \citep{baik2005phase,benaych2012singular}. Below the threshold the recovered singular vectors are asymptotically uncorrelated with the true ones; above it, let $\tilde u_i$ be the unit empirical left singular vector associated with spike $i$ and $u_i$ its true output direction. Their squared overlap $|\tilde u_i^{\top}u_i|^2$ is the squared cosine of their angle and converges to
\begin{equation}
  \rho(\gamma) = \frac{\gamma^4-q}{\gamma^4+q\,\gamma^2},
  \label{eq:bbp-overlap}
\end{equation}
rising from zero at threshold toward one \citep{benaych2012singular}. The corresponding input overlap $|\tilde v_i^{\top}v_i|^2$ converges to $(\gamma^4-q)/(\gamma^4+\gamma^2)$. The two coincide for a square matrix, $q=1$, the case used in Sec.~\ref{sec:phase}, where feature $i$ stops being resolvable at $T=\eta_i$. Here $T$ controls data noise and therefore the spectral resolution of a population feature. It differs from a dynamics-side temperature $T_{\mathrm{eff}}$ or $T_L$, which controls fluctuations of the parameters on a given landscape. Vanishing overlap means that the empirical singular vector is asymptotically uncorrelated with the population feature under these assumptions; it does not imply that every estimator fails or that the data Fisher matrix vanishes. The connection to Sec.~\ref{sec:breaking} is therefore an effective modelling choice: in Sec.~\ref{sec:phase} we use $\rho_i(T)$ as the weight of the corresponding feature error. This provides a soft counterpart of the exact filtration. It is not an identity between spectral overlap and Fisher information. For a uniform penalty, the ordered $\eta_i$ give the same ordering of the resolution and activation thresholds in this effective model; unequal penalty weights can make the orderings compete.

\begin{figure}[tbp]
  \centering
  \includegraphics[width=\textwidth]{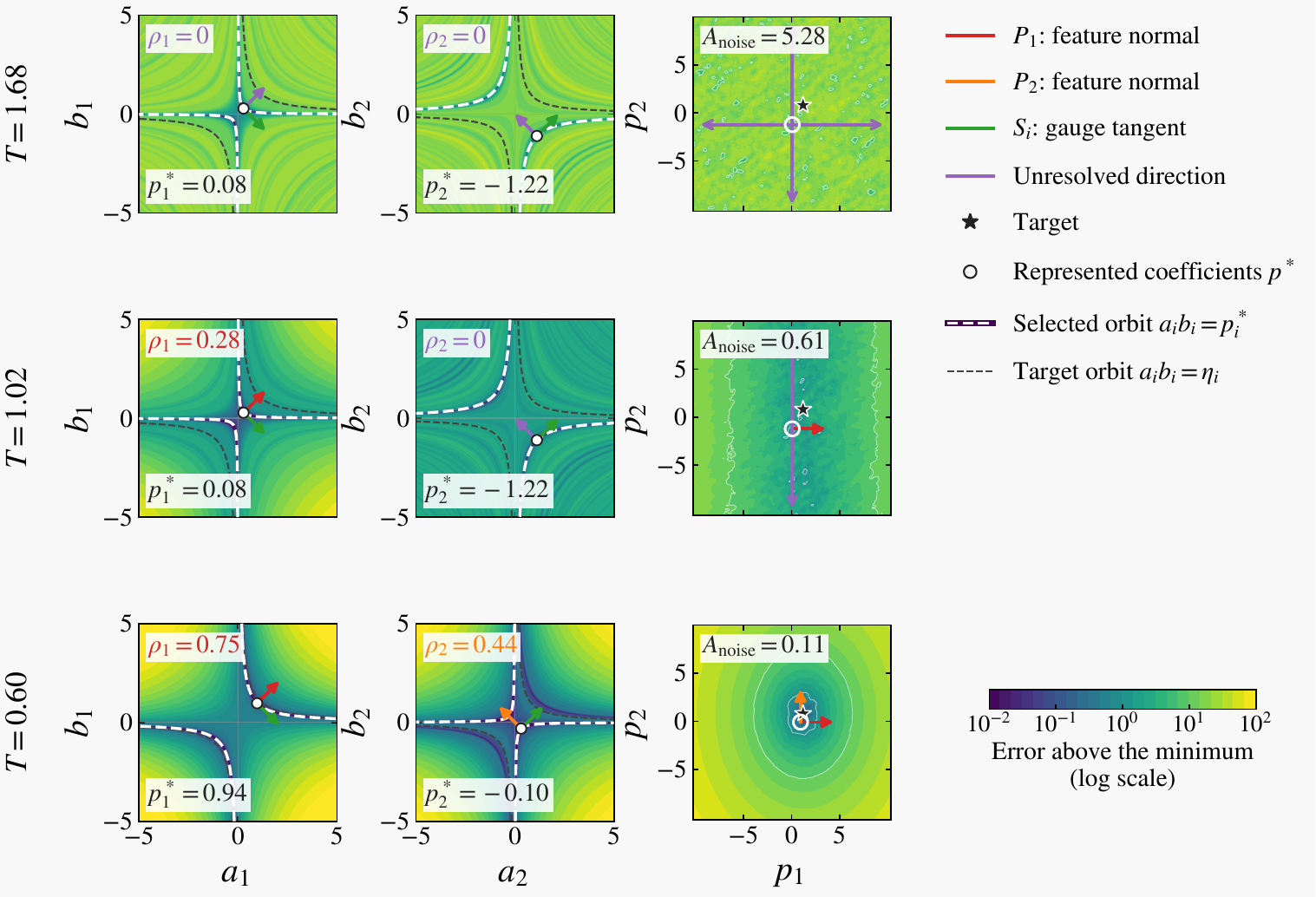} 
  \caption{\textbf{Illustrative landscapes at three levels of spectral resolution.} The target coefficients are $(\eta_1,\eta_2)=(1.2,0.8)$. For the square spiked-matrix model, the prescribed overlaps are $\rho_i(T)=(1-T^2/\eta_i^2)_+$. From top to bottom, $T=1.68$, $1.02$, and $0.60$, giving $(\rho_1,\rho_2)\approx(0,0)$, $(0.28,0)$, and $(0.75,0.44)$. Thus neither, only the first, and both population feature directions have nonzero asymptotic spectral overlap. The left and middle columns show the factor planes $(a_1,b_1)$ and $(a_2,b_2)$; the right column shows coefficient space $(p_1,p_2)$. The star marks the target, while the red-outlined circle marks the displayed configuration $p^*$. Its reported values are approximately $(0.08,-1.22)$ in the first two rows and $(0.94,-0.10)$ in the bottom row. White dots in the factor planes represent the corresponding coefficients. Black dashed hyperbolae mark $a_ib_i=\eta_i$; white curves mark $a_ib_i=p_i^*$. Green arrows indicate product-preserving tangents, and red or orange arrows indicate product-changing normals. These arrows identify geometric directions, without by themselves establishing Hessian eigenvalues at the marked configurations. All three columns use the same additional rough landscape, with amplitudes $A_{\mathrm{noise}}\approx5.28$, $0.61$, and $0.11$. This roughness is distinct from the deterministic overlap weights and is not generated by the BBP overlap formula alone. The resolution labels refer to singular-vector recovery, rather than exact fitting of the displayed coefficients. The random-field construction and grid selection of $p^*$ are specified in Appendix~\ref{app:numerics}.
}
  \label{fig:bbp}
\end{figure}

\subsection{The \texorpdfstring{$\beta$--$T$}{beta-T} phase diagram}
\label{sec:phase}
The parameter $T$ is the data-noise amplitude in Eq.~\eqref{eq:spiked}. A limited temperature analogy follows from the Gaussian likelihood, not from an identification with dynamical noise. For a data set $D$ of $N$ independent labels with noise variance $\sigma^2$ and mean half-squared empirical error $E_{\mathrm{emp}}$, Bayes' rule gives
\begin{equation}
 p(\theta\mid D)\propto\pi(\theta)
 \exp\!\left[-\frac{N E_{\mathrm{emp}}(\theta)}{\sigma^2}\right],
 \qquad \frac{\sigma^2}{N}=\frac{T^2}{n_{\mathrm{in}}},
 \label{eq:likelihood-scale}
\end{equation}
where $\pi(\theta)$ is a specified prior \citep{bishop}. This is a likelihood scale for the stated normalization. It neither derives the feature weighting below nor identifies $T$ with a Langevin temperature; in particular, it does not establish escape from a basin when a spectral overlap vanishes.

To illustrate competition between feature strength and regularization, we allow the two feature channels to carry different penalty weights $s_i>0$. The data strengths remain $\eta_i$; a larger $s_i$ simply makes that channel more costly to represent. We choose $s=(3,1)$ so that regularization penalizes the stronger feature more heavily, allowing the phase boundaries to cross. The ordinary $L_2$ penalty is recovered by setting both weights to one. The phase diagram uses the effective loss
\begin{equation}
 \mathcal L_{\beta,T}
 =\frac12\sum_{i=1}^2\rho_i(T)(p_i-\eta_i)^2
  +\frac\beta2\sum_{i=1}^2s_i(a_i^2+b_i^2),
 \qquad \rho_i(T)=\left(1-\frac{T^2}{\eta_i^2}\right)_+ .
 \label{eq:phase-loss}
\end{equation}
The targets remain $\eta_i$; the overlap weights the errors. Using it this way is an effective model of feature availability, not a simulation of noisy samples or a consequence of BBP alone. In particular, it differs from replacing the target by $\rho_i\eta_i$, although both choices give the same zero-activation threshold. Minimizing the penalty at fixed $p_i$ gives $\beta s_i|p_i|$. For $\rho_i>0$ the minimum is
\begin{equation}
 p_i^*=\left(\eta_i-\frac{\beta s_i}{\rho_i(T)}\right)_+ .
 \label{eq:phase-product}
\end{equation}
For $\rho_i=0$ and $\beta>0$ the minimum is $a_i=b_i=0$. At $\rho_i=\beta=0$, that factor pair is unconstrained. For $\beta>0$, feature $i$ has a nonzero minimizing coefficient exactly while
\begin{equation}
  \beta < \beta_c^{(i)}(T) = \rho_i(T)\,\frac{\eta_i}{s_i},
  \label{eq:phase-boundary}
\end{equation}
and the four regions have the feature-subset labels of the two filtrations that impose feature~1 first or feature~2 first (Sec.~\ref{sec:chain}). Their regularized minimizers do not generally satisfy the original exact constraints. At $T=0$ the thresholds are $\eta_i/s_i$: this is the feature-weighted extension of Sec.~\ref{sec:l2}, reducing to its thresholds when $s_i=1$. Distinct ratios give separate transitions. Raising $T$ lowers both thresholds through $\rho_i$ and bends the boundaries toward smaller $\beta$: a feature the data only partly resolves is relinquished sooner. Because $\rho_i$ depends on $\eta_i$ while the penalty weights $s_i$ do not, the two boundaries need not keep their order. They cross when $s_1/s_2>\eta_1/\eta_2$, here at $T\approx0.64$, $\beta\approx0.29$: below the crossing the penalty ordering governs and the heavily weighted feature is given up first, above it the resolution ordering governs and the weaker feature goes first.
For comparison with the deterministic activation boundaries, the numerical protocol uses overdamped Langevin dynamics, discretized as
\begin{equation}
 \theta_{n+1}=\theta_n-\Delta t\,\nabla\mathcal L_{\beta,T}(\theta_n)
       +\sqrt{2T_L\Delta t}\,\xi_n,
 \qquad \xi_n\sim\mathcal N(0,I),
 \label{eq:langevin-protocol}
\end{equation}
with independent Gaussian kicks at successive steps and $T_L=10^{-2}$ held fixed throughout the $(\beta,T)$ grid. The kicks perturb the parameters during optimization, whereas $T$ changes the prescribed error weights $\rho_i(T)$. They are part of the numerical optimization protocol and do not implement the data noise represented by $T$. Here ``Langevin sampling'' means collecting configurations from a finite-time trajectory; equilibration is not established by the protocol alone. No noisy data matrices $Z$ are generated, so the numerical panel does not average over data-noise realizations.

The feature classification also has a separate threshold. For a fraction $0<\alpha<1$, the deterministic minimum obeys
\begin{equation}
 p_i^*>\alpha\eta_i
 \quad\Longleftrightarrow\quad
 \beta<(1-\alpha)\,\beta_c^{(i)}(T),
 \qquad \rho_i(T)>0.
 \label{eq:classification-boundary}
\end{equation}
The implemented choice $\alpha=0.2$ therefore lowers the classification boundary to $0.8\beta_c^{(i)}(T)$ even before finite-time or noise effects. Changing a common $\alpha$ rescales both deterministic curves by the same factor: the crossing value of $T$ is unchanged and its $\beta$ coordinate is multiplied by $1-\alpha$. This prediction for the minimum does not determine the threshold dependence of a finite-noise time average.
\begin{figure}[H]
  \centering
  \IfFileExists{figures/phase_diagram.pdf}{%
    \includegraphics[width=\textwidth]{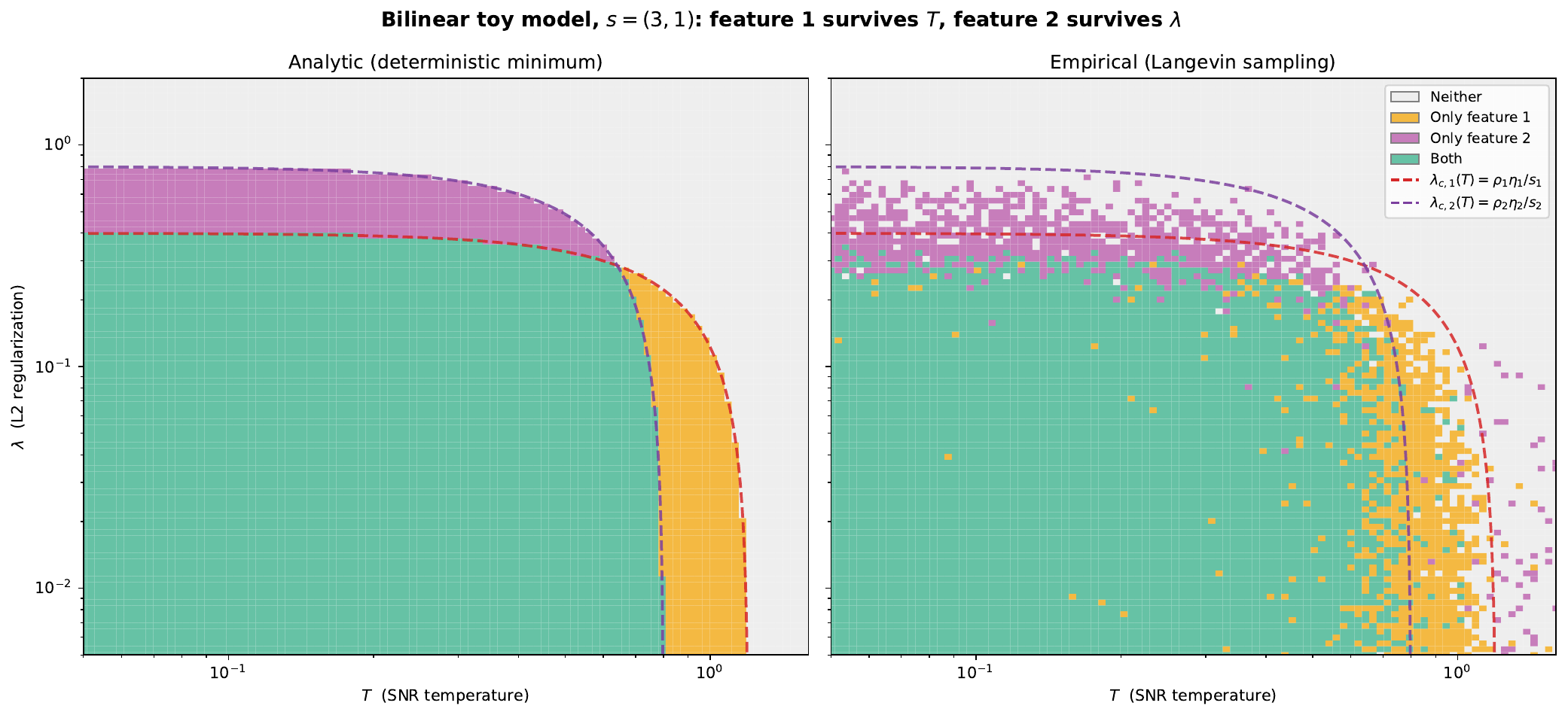}}{%
    \fbox{\parbox{\textwidth}{\centering\todo{Insert the $\beta$-$T$ phase diagram as figures/phase\_diagram.pdf.}}}}
  \caption{ \textbf{Feature selection under regularization and prescribed spectral-resolution weights.} Two-feature bilinear model $p_i=a_ib_i$, with targets $\eta=(1.2,0.8)$ and penalty weights $s=(3,1)$. Both panels vary $\beta$ and $T$ in the effective loss $\mathcal L_{\beta,T}$ of Eq.~\eqref{eq:phase-loss}. Colour identifies the feature subset satisfying the criterion used in each panel: neither feature (grey), only feature~1 (orange), only feature~2 (purple), or both (green). \textbf{(a)} Analytic classification of the global minimum. Feature $i$ is active when $p_i^*>0$, equivalently $\beta<\beta_c^{(i)}(T)=\rho_i(T)\eta_i/s_i$. The panel is constructed directly from these inequalities. The dashed curves show the activation boundaries and are repeated in panel~(b). \textbf{(b)} Classification from finite-time Langevin trajectories on the same effective loss, with fixed $T_L=10^{-2}$, $\Delta t=10^{-3}$, and $1.2\times10^4$ steps. Each cell starts from $\theta\sim\mathcal N(0,0.1^2)$. After the initial $40\%$ of the run, products $p_i$ are recorded every 20 steps; feature $i$ is classified as active when their time average exceeds $0.2\eta_i$. There is one trajectory per cell on a $100\times100$ grid. The Gaussian kicks perturb parameters; the resolution weights are prescribed, and no noisy datasets are sampled. The analytic activation boundaries cross at $T\approx0.64$, $\beta\approx0.29$. The respective weights vanish at $T=0.8$ and $T=1.2$. The panels use different activity criteria: even at the deterministic minimum, the criterion $p_i^*>0.2\eta_i$ gives the boundary $\beta=0.8\beta_c^{(i)}(T)$. Parameter noise and finite relaxation time can cause further differences; the numerical panel does not establish equilibration. Grid ranges, sampling indices, and random-number settings are given in Appendix~\ref{app:phase}.}
  \label{fig:phase}
\end{figure}


\section{Deep linear networks}


\label{sec:where}
We now allow full matrix factors. A depth-$N$ linear network has $N-1$ hidden layers and realizes $f_\theta(x)=Mx$ with $M=W_N\cdots W_1$; a two-factor network therefore has one hidden layer. We leave the diagonal minimal model for these linear and deep linear networks, where the contingent group can still be written down (Sec.~\ref{sec:linear}),
for one hidden layer the regularized Hessian admits the four-sector decomposition of Sec.~\ref{sec:sectors}, and the regularization path selects feature subsets that can label a nested constraint filtration (Sec.~\ref{sec:nested}). Nonlinear networks and other learning systems are discussed in Sec.~\ref{sec:discussion}.

\subsection{The contingent group of linear networks}
\label{sec:linear}
\begin{figure}[H]
  \centering
  \includegraphics[width=0.7\textwidth]{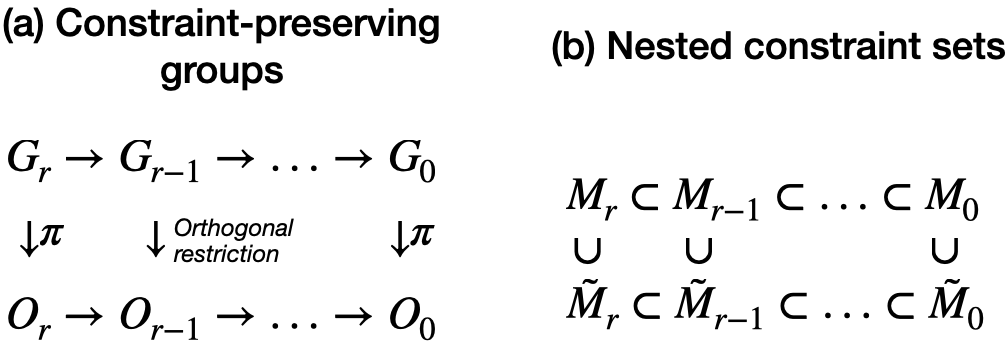}
  \caption[Constraint hierarchies and their norm-preserving restrictions]{%
  \textbf{Constraint hierarchies and their norm-preserving restrictions.}
  \textbf{(a)}~\emph{Upper row:} the groups $G_k$, where $G_k$ preserves each of the first $k$ coefficient functionals for every represented map.
  Reading from left to right removes constraints and enlarges the group, $G_r\subseteq G_{r-1}\subseteq\cdots\subseteq G_0$; reading from right to left adds constraints.
  \emph{Lower row:} their norm-preserving restrictions $O_k=G_k\cap[\OG(n_{\mathrm{out}})\times\OG(n_{\mathrm{in}})]$.
  \emph{Vertical arrows:} restriction to norm-preserving transformations, not a group projection.
  These restrictions preserve the specified coefficients; preservation of the full data error is a separate condition.
  \textbf{(b)}~\emph{Upper row:} the exact constraint sets $\mathcal M_k=\{\theta:c_i(\theta)=\eta_i,\ i\leq k\}$, with $\mathcal M_0=\Theta$.
  Each additional equality selects a subset of the preceding set.
  \emph{Lower row:} the balanced subsets $\widetilde{\mathcal M}_k=\{(W_2,W_1)\in\mathcal M_k:W_2^{\top}W_2=W_1W_1^{\top}\}$.
  Both rows are therefore nested, and $\widetilde{\mathcal M}_k\subseteq\mathcal M_k$.
  A constraint set can contain multiple represented maps, matrix ranks, and symmetry orbits.
  Balancedness selects minimum-norm factorizations at a fixed map; it does not reduce the constraint set to a single configuration or identify it with a regularized minimum.}
  \label{fig:groups}
\end{figure}

In linear and deep linear networks the contingent group can be written down explicitly. With the features the singular pairs $(u_i,v_i)$ of $\Syx$, their coefficients $c_i(M)=u_i^{\top}Mv_i$, and a pair $(g,h)$ acting on the end-to-end map by $M\mapsto gMh^{\top}$, Eq.~\eqref{eq:filtration-group} becomes
\begin{equation}
  \begin{aligned}
  &G_d=\bigl\{(g,h)\in\GL(n_{\mathrm{out}})\times\GL(n_{\mathrm{in}}):\quad\\
  &\qquad u_i^{\top}gMh^{\top}v_i=u_i^{\top}Mv_i\ \text{for all }M\text{ and }i\le d\bigr\}.
  \end{aligned}
  \label{eq:Gk}
\end{equation}
Its elements rescale each constrained pair, $g^{\top}u_i=\lambda_iu_i$ and $h^{\top}v_i=\lambda_i^{-1}v_i$, act freely on the two unlearned complements, and may map learned directions into them (App.~\ref{app:group}), so $G_d$ lies in the parabolic subgroup that preserves the complements, and its dimension
\begin{equation}
  \dim G_d=d+(n_{\mathrm{out}}-d)\,n_{\mathrm{out}}+(n_{\mathrm{in}}-d)\,n_{\mathrm{in}}
  \label{eq:dimGk}
\end{equation}
falls with each added scalar constraint. Here $d$ counts imposed coefficients; these constraints alone do not require $\rank M=d$. The group preserves the coefficient functionals, rather than fixing one particular matrix, and a solution set need not be a single orbit. A point stabilizer would instead be the subgroup leaving one specified configuration unchanged; we do not use that name for $G_d$. Figure~\ref{fig:groups} summarizes the constraint hierarchy and separates it from the norm-preserving architectural action.

\subsection{Regularization introduces additional massive modes}
\label{sec:sectors}
The distinction between architectural redundancy and unresolved features helps interpret the Hessian spectrum of the regularized linear network; the eigenvalues themselves require the loss calculation. \citet{cascade} derived this spectrum in closed form and sorted it into four classes: massive data-dependent eigenvalues, a null-space sector, a $\beta$ sector from the explicitly broken symmetry, and zero eigenvalues from symmetry. For one hidden layer the eigenvalues in their Table~II fall into the four sectors below; their loss, $\lVert\cdot\rVert^2+\beta\lVert\theta\rVert^2$, is twice ours, so their eigenvalues are twice those quoted here. A first pass through the symmetry classification suggests only two families: flat modes from the surviving generators, massive modes from the broken ones. The distinction between nonzero symmetry-orbit tangents and perturbations that change the product only at higher order supplies the additional separation, as the onset calculation below shows. In this section, $E$ in Eq.~\eqref{eq:regloss} denotes the squared error $\tfrac12\lVert W_2W_1-\Syx\rVert_F^2$ for whitened inputs and input-output cross-covariance $\Syx$; unlike $E_r$, it fixes every coefficient of the map, not only its $r$ coefficients along the features. Before setting them out, one point of bookkeeping. The symmetries of Sec.~\ref{sec:symmetry} were derived from the map $M$ and make no reference to $\beta$; the sectors below are sectors of $\nabla^2\mathcal{L}_\beta$, Eq.~\eqref{eq:regloss}, at a balanced rank-$d$ critical point of $\mathcal{L}_\beta$, expressed in an aligned hidden basis. Here balanced means $W_2^{\top}W_2=W_1W_1^{\top}$; in this basis $d$ factor pairs are nonzero and the remaining hidden-unit blocks vanish. The explicit sector formulas below assume one hidden layer, $\beta>0$, and generic distinct signal strengths; positivity of exchange modes additionally requires a stable selected feature subset. The two statements are compatible but not interchangeable, because the penalty is not invariant under everything the map is. Of the architectural group \eqref{eq:internal}, only the orthogonal subgroup preserves $\lVert\theta\rVert^2$, so only $\OG(w)$ leaves $\mathcal{L}_\beta$ invariant while all of $\GL(w)$ leaves $E$ invariant; the difference is what produces the pseudo-Goldstone sector. The same restriction applies to the contingent action $(W_2,W_1)\mapsto(gW_2,\,W_1h^{\top})$, which preserves the Frobenius norms only for orthogonal $g,h$.

\begin{itemize}
  \item \textbf{Flat modes are tangents to architectural symmetry orbits.} At a stationary point, each nonzero tangent induced by a continuous symmetry of the objective is a null direction of $H_{\mathcal L}$; linearly independent tangents, rather than abstract generators, determine the count. At a rank-$d$ critical point, $d$ hidden units are active and the other $w-d$ are inactive, with zero weights. Orthogonal reparametrizations \eqref{eq:internal} that rotate the active units, among themselves or into the inactive ones, change the weights but neither the end-to-end map nor the norm, so they move the configuration along an equal-loss family of critical points (minima when the point is stable); these are the flat directions, $\dim\OG(w)-\dim\OG(w-d)$ of them (App.~\ref{app:goldstone}). In this one-hidden-layer setting, learning one more independent feature activates one more unit and reduces the point stabilizer from $\OG(w-d)$ to $\OG(w-d-1)$, the matrix counterpart of the sign symmetry of Sec.~\ref{sec:l2}. Rotating only the inactive units has no effect at all, since their weights are zero. Away from degeneracies and thresholds, the listed contingent generators supply no additional Hessian-null directions (Appendix~\ref{app:group}): those acting on the unlearned complements touch only weight blocks that are zero at the critical point, the rescalings of the learned pairs act inside the architectural orbit, and those that mix a learned direction into a complement change the end-to-end map and are Fisher-positive; their Hessian curvature is determined by the stability of the critical point.
  \item \textbf{Onset modes activate an inactive feature at second order.} Choose an inactive hidden unit and an unrepresented target direction $(u_b,v_b)$. Perturb that unit's input weights by $\varepsilon v_b^{\top}$ and its output weights by $\delta u_b$. Since both blocks initially vanish, the first-order product differential is zero, while the exact product change is $\Delta M=\varepsilon\delta\,u_bv_b^{\top}$. For the full two-factor objective, the resulting change is
  \begin{equation}
    \Delta\mathcal L_\beta
       =\frac\beta2(\varepsilon^2+\delta^2)
          -\eta_b\varepsilon\delta
          +\frac12\varepsilon^2\delta^2,
    \label{eq:onset-expansion}
  \end{equation}
  so the Hessian in $(\varepsilon,\delta)$ is
  \begin{equation}
    \begin{pmatrix}\beta&-\eta_b\\-\eta_b&\beta\end{pmatrix},
    \qquad \lambda_\pm=\beta\pm\eta_b.
    \label{eq:onset-hessian}
  \end{equation}
  Fisher information vanishes on these directions at the inactive configuration because $D M=0$. The $\beta$ contribution is the penalty curvature and the $\pm\eta_b$ contribution comes from the data-error residual. The smaller eigenvalue crosses zero at $\beta=\eta_b$. These are onset directions, not nonzero tangents of a symmetry orbit: a homogeneous action on the zero blocks leaves them zero, whereas the perturbations above create a new product. Their link to the filtration is that they change only $c_b$, which a partial task omitting feature $b$ leaves unconstrained. The eigenvalues follow from the full objective, not from this freedom alone. For a fully inactive path through more than two weight matrices, the product changes at order greater than two; its quadratic curvature is then only the penalty, and first-order global onset need not have a Hessian zero crossing (Sec.~\ref{sec:nested}).
  \item \textbf{Function-changing modes have positive Fisher norm.} With whitened inputs and a positive-definite Gaussian output metric, a first-order change of the end-to-end map has positive Fisher norm. At a regularized point, however, the error Hessian also contains a residual term, so its eigenvalues must be computed separately. One cannot simply declare the continuous MSE symmetry trivial: $(g,h)=(cI,c^{-1}I)$ acts identically on every map, and orthogonal rotations in true null-signal subspaces can also survive. Identifiability is determined by the product differential after such redundancies are accounted for. Among the function-changing directions are the exchange directions of \citet{cascade}, which tilt a learned mode $a$ toward an unlearned mode $b$; with one hidden layer and whitened inputs their eigenvalues are $\eta_a\pm\eta_b$.
  \item \textbf{$\beta$-modes are pseudo-Goldstone.} The $L_2$ term reduces the reparametrization group from $\GL$ to its orthogonal subgroup, keeping only the norm-preserving maps. At the balanced critical point, the architectural tangent space splits orthogonally into tangents induced by skew-symmetric and symmetric hidden-coordinate generators. Their $H_{\mathcal L}$ eigenvalues are respectively $0$ and $2\beta$, and both sectors are Fisher-null (Appendix~\ref{app:goldstone}). Generators supported entirely on inactive hidden units induce zero tangents and are excluded from the count.
  Linear scaling with $\beta$ supports this interpretation only together with gauge-orbit alignment and vanishing Fisher norm.
\end{itemize}

The shift has a consequence for the flat sector that is worth stating, since it is easy to read ``flat'' as ``no curvature in the error''. It is not. Let $v\ne0$ be tangent to the $\OG(w)$ orbit at a critical point $\theta^{*}$ of $\mathcal{L}_\beta$. Both $E$ and $\lVert\theta\rVert^2$ are constant along that orbit, so $\mathcal{L}_\beta$ is too and $\nabla^2\mathcal{L}_\beta(v,v)=0$. By Eq.~\eqref{eq:regloss},
\begin{equation}
  \nabla^2 E(v,v) \;=\; -\,\beta\,\lVert v\rVert^2 \;<\; 0 .
  \label{eq:flat-is-concave}
\end{equation}
The exactly flat directions of the regularized loss are therefore \emph{concave} directions of the unregularized error, with curvature exactly $-\beta$. Flatness here is a cancellation between a concave error surface and the penalty, not an absence of curvature in $E$; the Goldstone modes of the broken symmetry sit precisely where the error landscape curves downward, which is where the concave boundaries of the $L_2$ mechanism are found \citep{ersoy2}. The value $-\beta$ itself is automatic, since $\nabla^2\mathcal{L}_\beta=\nabla^2E+\beta I$ holds everywhere; what the symmetry argument predicts is the number of these directions, $\dim\OG(w)-\dim\OG(w-d)$ under the stated one-hidden-layer assumptions, for $\beta>0$ away from transitions and accidental degeneracies, and that they are the orbit directions of $\OG(w)$. In short: the Hessian at a rank-$d$ critical point resolves into four sectors: flat, onset, massive, and pseudo-Goldstone, and all four are organized by the symmetries classified in Sec.~\ref{sec:threetypes}: exactly flat where an architectural generator survives the penalty but is broken spontaneously by the configuration, onset where the contingent freedom of an unlearned feature is realized at second order, massive where the data has already locked one in, and pseudo-Goldstone where the regularizing field has explicitly broken an architectural generator. In the terms of \citet{cascade} these are, in turn, their zero eigenvalues, null-space sector, massive data-dependent sector and $\beta$ sector. They trace the first to the orthogonal symmetry and the last to the explicit breaking of $\GL(w)$ by the penalty; what the present account adds is the origin of the other two in the contingent freedom of the partial task, and which of the four carries information. The comparison therefore distinguishes two effects: informative task curvature and explicit gauge lifting. An onset soft mode is not itself a Goldstone mode, and a higher-order onset direction must not be identified with a nonzero symmetry-orbit tangent. The first is the introduction of a feature (Sec.~\ref{sec:emergent}): the mass corresponds to a nonzero Fisher eigenvalue, and the direction is an emergent learning direction. Along the one-hidden-layer cascade, the onset direction becomes soft at $\beta_c$ and the active branch then has a function-changing, Fisher-positive normal direction. This is the counterpart of the emergent learning direction in the filtration, under a different comparison of states. At the collapsed threshold $\beta=\eta_b$, the Fisher metric still vanishes in that direction. It becomes positive on the nonzero branch for $\beta<\eta_b$. The onset eigenvalue is not a Fisher eigenvalue. The Fisher metric vanishes along the onset direction until the feature is learned, so $\beta\pm\eta_b$ comes entirely from the penalty ($\beta$) and from the residual of the still-unlearned feature ($\pm\eta_b$). The vanishing is forced by symmetry: the inactive unit sits at the fixed point of the sign flip of its weights, an architectural symmetry, and at such a point the Fisher information is singular along the flipped directions \citep{ziyin-sym}. The second is the introduction of the regularizer, which produces the pseudo-Goldstone $\beta$-modes. They lie along the architectural orbit, on which the realized function does not change, so they remain null directions of the Fisher metric; their total curvature is $2\beta$, comprising $\beta$ from the ordinary error Hessian and $\beta$ from the penalty Hessian at the regularized critical point. Although its value is controlled by $\beta$, this curvature does not imply statistical distinguishability: these gauge directions remain Fisher-null. In this setting, each activation changes the locally identifiable parameter subspace. For the full two-factor matrix model at a canonical balanced rank-$d$ factorization, $\rank\Fisher=d(n_{\mathrm{out}}+n_{\mathrm{in}}-d)$ under the positive observation metric. A rank increment can therefore create several Fisher-positive directions, including mixing directions. The one-mode-per-constraint identity of Sec.~\ref{sec:rankone} concerns one added scalar observation at a common regular point, not every rank transition of a full network.

\subsection{Nested constraints and ordered regularized phases}
\label{sec:nested}
\begin{figure}[tbp]
  \centering
  \includegraphics[width=0.66\textwidth]{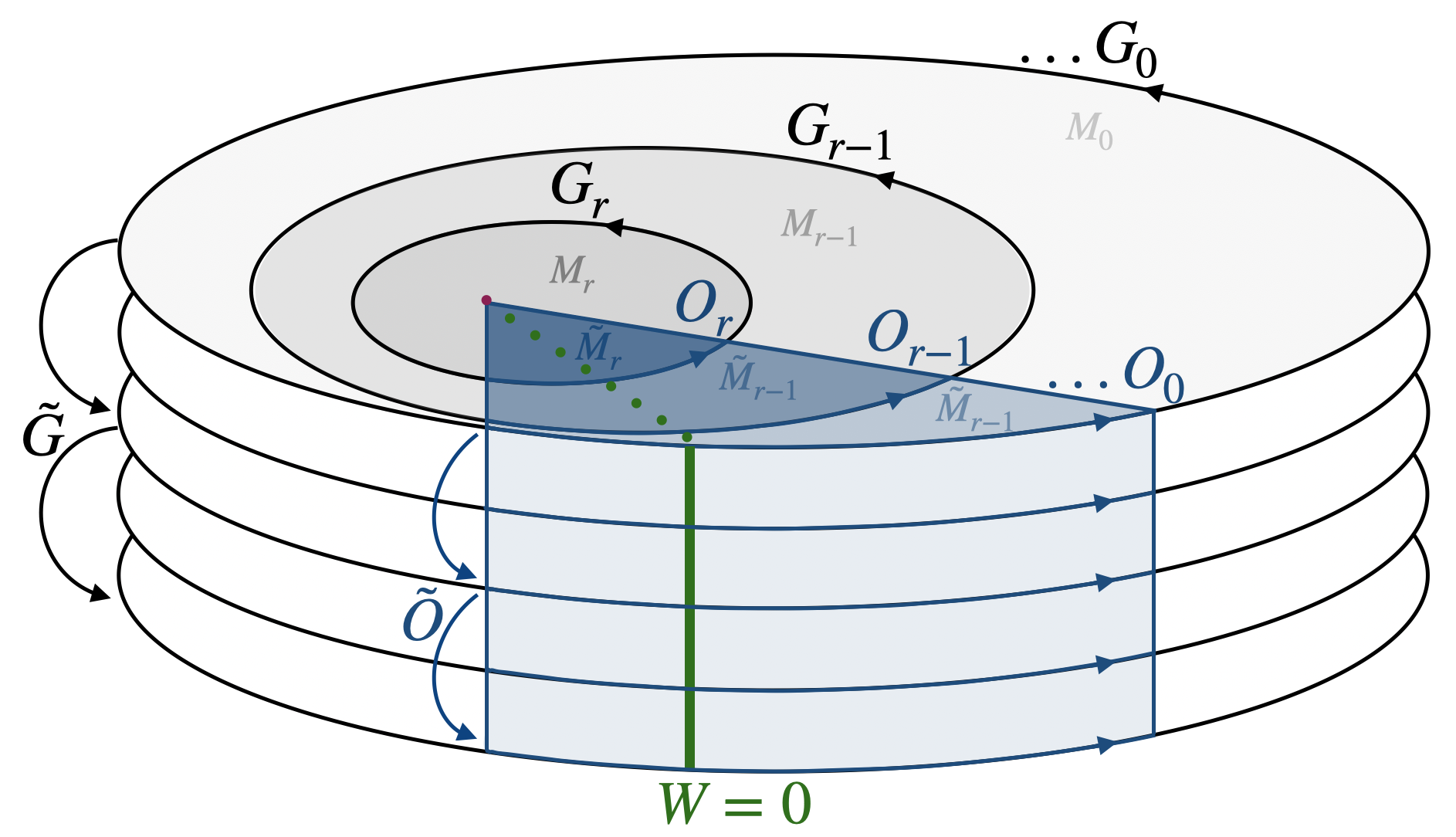}
  \caption[Nested feature constraints and the remaining freedom of representation]{%
  \textbf{Nested feature constraints and the remaining freedom of representation.} Schematic for a one-hidden-layer linear network $M=W_2W_1$ of hidden layer width $w$, with a fixed ordering of the features.
  \emph{Grey regions:} the constraint sets $\mathcal M_k$ of Eq.~\eqref{eq:filtration-minima}, nested from $\mathcal M_0=\Theta$ (outermost, unconstrained) to $\mathcal M_r$ (innermost, all $r$ constraints). Each step inward adds one constraint and keeps the earlier ones. $\mathcal M_r$ is a set of exact solutions, not a single configuration.
  \emph{Blue shading:} the balanced subsets $\widetilde{\mathcal M}_k=\mathcal M_k\cap\mathcal B$, where $\mathcal B$ is the set of balanced factorizations, $W_2^{\top}W_2=W_1W_1^{\top}$.
  \emph{Arrows:} the black arrows $G_k$ are transformations that preserve the first $k$ coefficient functionals and act within $\mathcal M_k$, which need not be a single orbit.
  The blue arrows $O_k$ are the norm-preserving parts of these contingent actions. They keep both balancedness and the imposed coefficients.
  \emph{Vertical copies:} factorizations of the same map $M$, related by $\widetilde G=\GL(w)$ acting as $(W_2,W_1)\mapsto(W_2G,G^{-1}W_1)$ (side arrows).
  Its orthogonal subgroup $\widetilde O=\OG(w)$ also preserves the parameter norm and balancedness.
  The copies show architectural redundancy, not further constraints.
  \emph{Green dotted path:} connects representative configurations as constraints are added and ends at the balanced exact solution $\theta^{*}\in\widetilde{\mathcal M}_r$ (magenta).
  The path only illustrates the filtration. It is neither a computed training trajectory nor a path of regularized minimizers; regularized minimizers can pass through the same ordered feature subsets with coefficients shifted away from the exact targets.
  The zero configuration lies in $\mathcal M_0$ but satisfies none of the constraints, since every $\eta_i>0$.
  Areas, distances, and vertical spacing are schematic, and the regions do not depict basins of attraction.}
  \label{fig:orbits}
\end{figure}

For a fixed ordering and retained target values, the partial-task solution sets are nested. Separately, in the aligned separable model the regularization path selects feature subsets in an ordered sequence. This is what Figs.~\ref{fig:groups} and \ref{fig:orbits} depict; here we argue it must be so.

\begin{claim}
Consider the aligned separable deep-linear model with distinct singular values $\eta_1>\dots>\eta_r>0$, uniform layerwise $L_2$ regularization, and every hidden width at least $r$. Then the solution sets of the partial tasks form a nested family $\mathcal{M}_0\supset\mathcal{M}_1\supset\dots\supset\mathcal{M}_{r}$, and the quasi-static regularization path from the zero map toward the rank-$r$ target map passes through phases corresponding to them, in order and one transition per step: its rank-$k$ phase represents the first $k$ features. Only with one hidden layer is the active coefficient exactly $\eta_i-\beta$; at greater depth it follows the corresponding nonlinear stationary equation. The correspondence concerns feature subsets, not membership of the regularized minimizer in the exact sets $\mathcal M_k$.
\end{claim}

\noindent The argument separates constraint-set inclusion, global threshold ordering, and local stability.

\begin{enumerate}
  \item \textbf{Containment.} A solution that fits the first $k{+}1$ singular directions also fits the first $k$. Hence the solution sets are nested, $\mathcal{M}_{k+1}\subset\mathcal{M}_{k}$, and the symmetry groups are nested in the same order, \eqref{eq:chain}. In the minimal model the modes decouple and thus one could think of differently ordered nesting. The one given here is the one taken by the investigated $L_2$ path.
  \item \textbf{One feature at a time.} Because the modes decouple, each singular direction has its own critical strength $\beta_c^{(i)}$, and, for a uniform penalty, the ordering of the $\eta_i$ orders the $\beta_c^{(i)}$. As $\beta$ is lowered the directions turn on one at a time, in descending order: rank increases in unit steps, never by two at once for generic $\eta_i$. Established: \citet{cascade}; the analogous order in time is the classic deep-linear result of \citet{saxe}.
  \item \textbf{The quasi-static path cannot pass early.} At the rank-$k$ critical point every onset direction toward an unlearned feature has positive curvature while $\beta>\beta_c^{(k+1)}$ (with one hidden layer it is $\beta-\eta_b$; with more, the penalty $\beta$ alone), and the exchange directions are positive as well. The rank-$k$ configuration is therefore a local minimum of $\mathcal{L}_\beta$, and comparison of the decoupled objective values, rather than local curvature alone, determines when it ceases to be the global minimum, so the quasi-static path of Sec.~\ref{sec:l2}, which tracks the minimum at every $\beta$, cannot leave it earlier. With one hidden layer the onset is continuous, and $\beta_c^{(k+1)}$ is also where the rank-$k$ configuration loses stability. With two or more hidden layers the stationarity condition is nonlinear and the two configurations coexist over a finite range of $\beta$ \citep{ziyin-ueda,grokking}: the rank-$k$ configuration stays locally stable below $\beta_c^{(k+1)}$, and the path leaves it there because the rank-$(k{+}1)$ configuration becomes lower. This is established for the swept path; what it does \emph{not} establish is that the sublevel sets of $\mathcal{L}_\beta$ are disconnected between ranks, nor that arbitrary dynamics follow the same sequence. Those questions require separate analysis of barriers and attraction basins. For a specified deterministic training dynamics, an attraction basin is the set of initial configurations converging to a given minimum or minimum set; it is not the exact constraint set $\mathcal M_k$.
\end{enumerate}

\noindent Constraint retention gives the nesting. The distinct global activation thresholds of the separable model give the ordered feature subsets along a continuous sweep of global minimizers. Local-minimum continuation can instead show metastability and hysteresis; neither path establishes nested attraction basins.

\section{Discussion}

\label{sec:discussion}
\paragraph{Summary}
A partially specified learning task leaves contingent freedom; where a continuous action generates that freedom, adding a constraint reduces the corresponding symmetry. We classified the symmetries acting on such a system into three kinds, identified the contingent one, solved for it in deep linear networks, where it lies in a parabolic subgroup and its dimension falls with each coefficient constraint imposed, and distinguished its nontrivial action from transformations that are architectural or fix the configuration. Data distinguish the directions that change the observed map to first order. The generator actions and the loss expansion together organize the four curvature sectors found by \citet{cascade}: at a critical point the contingent generators act trivially on the weights, act inside the architectural orbit, or are locked by the data into massive directions, so their fingerprint is the structure of the onset sector rather than an extra flat direction. At a common regular exact fit, adding one independent scalar constraint increases both the error-Hessian rank and the Fisher rank by one. Along the regularization path, the related gain of statistical distinguishability is determined by the product Jacobian on the active branch; the exact-fit Hessian-Fisher identity does not apply to a regularized minimum with nonzero residuals (Appendix~\ref{app:goldstone}). Two measurable phenomena trace the breaking without requiring the group: a regularization sweep locates feature-selection events, and a spectral threshold characterizes recoverability in the specified noisy-data model.

\paragraph{Refining the task description.} The coarse-graining interpretation separates redundancy in representation from unresolved functional distinctions. Architectural freedom changes how a function is represented; contingent freedom can change aspects of that function omitted by a partial task. Adding an independent scalar feature refines the description and, at a common regular exact solution, increases both the error-Hessian rank and the Fisher rank by one. For the independent Gaussian observations used here, the reverse operation has a direct statistical meaning: discarding the additional observation recovers the preceding partial model and removes its Fisher contribution. This is a concrete instance of the monotonicity of Fisher information under parameter-independent processing of observations \citep{ay2015}. The statistical statement concerns the observation models; the deterministic maps in Fig.~\ref{fig:coarse-graining} organize the corresponding equivalences between parameters, functions, and retained feature values.
This interpretation remains meaningful when the fibers are not single symmetry orbits. It describes a hierarchy of task observables, without constructing a spatial rescaling or renormalization-group flow. The regularization sweep provides a separate comparison of fitted states under the full task; it does not itself implement the coarse-graining maps.

\paragraph{Nonlinear networks: what survives without a group description.}

\label{sec:survives}
\label{sec:neither}
The argument of the paper can now be stated in its general form. The constraint-based account of Sec.~\ref{sec:symmetry} applies more broadly than linear activations, but redundancy and incompleteness do not guarantee a continuous group action. The regularity and observation assumptions of Sec.~\ref{sec:setup} are needed for the local rank identities even when no group is available. Without the linear-model structure, three explicit descriptions are lost: the closed form \eqref{eq:dimGk}, the sector dimensions of App.~\ref{app:goldstone}, and the identification of which generator is consumed at which transition. What is not lost is the phenomenology of Sec.~\ref{sec:detecting}. A cascade of transitions under a swept $\beta$ remains observable, it is observed in nonlinear networks \citep{ersoy1,ersoy2}, but identifying a transition with subgroup reduction requires establishing the corresponding action. Likewise, a spectral threshold is informative only when the data model and estimator meet its assumptions. The two probes are complementary: the cascade measures which features survive a penalty, whereas the spectral calculation describes recoverability in a specified noisy ensemble. Their quantitative relation must be tested for the model at hand. A consequence follows, and it concerns what they do \emph{not} need. 
Neither measurement requires an explicit symmetry-group description, but the analytic spectral thresholds still require a suitable noise ensemble, scaling limit, and estimator. A regularization sweep and joint feature/Jacobian/Hessian measurements can be performed in nonlinear networks without assuming that their features are singular directions of an input-output covariance.

\paragraph{Beyond neural networks.}
\label{sec:universal}
The regular scalar-constraint identities of Sec.~\ref{sec:rankone} do not require linear activations, although the explicit toy groups and full-network spectra do use the stated model structure. The ingredients that drive the mechanism are two: a parameter space large enough to carry reparametrization redundancy, and unlearned directions along which a change can be absorbed into either the data or the parameters. Their realization depends on both the architecture and the task. Where the relevant unresolved variations can be identified, we expect a constraint-based description, and their breaking should produce the same signature: a Hessian that separates into a few massive, task-aligned modes and a large complement of flat ones, with the massive set growing each time a feature is learned. Related reorganizations of response spectra have been documented in physical learning systems, although their curvature operators need not be parameter Fisher matrices. Stern and coauthors show that learning imprints structure on the Hessian: the effective dimension of the response drops, the low-eigenvalue eigenvectors rotate to align with the task, and the cost and physical Hessians become coupled \citep{stern-effects,stern-become}; supervised learning in such networks was framed as machine learning becoming ``learning machines'' with exactly this stiff/soft organization \citep{stern-supervised}. The relation of those stiff and soft response modes to informative parameter directions is a hypothesis. Calling a soft mode Goldstone additionally requires a nonzero tangent generated by an exact continuous symmetry.

\paragraph{Observables.} 
The account predicts four things that can be checked. Turning on the $L_2$ field breaks $\GL\to\OG$ and, through the balance condition, selects a single $\OG(w)$-orbit of balanced states within each fiber, so positive regularization selects balanced representatives from the degenerate fiber. This does not force a discontinuity of a chosen minimizer at $\beta=0$: a balanced branch can approach a balanced member of the unregularized minimum set continuously. With one hidden layer, the Hessian at a rank-$d$ critical point should carry four eigenvalue classes with the dimensions of App.~\ref{app:goldstone}: flat, $2\beta$, $\beta\pm\eta_b$ (and $\beta$ along directions without signal), and data-massive, including $\eta_a\pm\eta_b$ along the exchange directions. For the stated one-hidden-layer linear model this is the spectrum derived by \citet{cascade}; analogous distinctions can guide nonlinear experiments, but the four exact classes and their multiplicities are not universal predictions. For one hidden layer, $\beta>0$, and the generic aligned configurations considered above, the symmetry-generated null space has dimension $\dim\OG(w)-\dim\OG(w-d)$ and is spanned by the orbit directions of $\OG(w)$; extra degeneracies must be checked separately. Along these directions the unregularized error is concave with curvature $-\beta$, Eq.~\eqref{eq:flat-is-concave}: an algebraic consequence of $H_{\mathcal L}=H_E+\beta I$. Their orbit alignment and dimension, rather than the value $-\beta$ alone, test the symmetry interpretation. The groups in \eqref{eq:dimGk} have decreasing dimension, but orbit dimensions also depend on the subgroup that leaves the chosen configuration unchanged (its point stabilizer); group dimension alone does not determine phase ordering or stability. Outside the solvable case, the informative test is to track the error on each feature, the data Jacobian or expected Fisher matrix, and both the error and penalized Hessians through a transition. This can establish whether that transition has the proposed mechanism. The general constraint principle alone does not require every architecture to exhibit a discrete $L_2$ cascade or a covariance-based BBP threshold.

\paragraph{Relation to symmetry-constraint accounts.}
The contingent symmetry is related to, but distinct from, a broader phenomenon: mirror symmetries can define invariant fixed subspaces whose occupation is favored by sufficiently strong regularization or gradient noise under the hypotheses of \citet{ziyin-sym}. A fixed subspace need not consist entirely of stationary points. That account concerns symmetries of the loss itself; it includes architectural examples such as hidden-unit permutations, with additional assumptions for the regularization and noise conclusions. Contingent symmetries here preserve the auxiliary partial loss, but need not preserve the full loss: they are the freedoms a not-yet-fully-specified \emph{task} leaves open, they form a nested chain indexed by the retained constraints. In the minimal model, their reduction under an independent scalar constraint is tied to an increase of Fisher rank at a common regular exact solution. The regularization path provides a related but distinct comparison.

\paragraph{Shortcomings.}
Two questions remain open. Section~\ref{sec:nested} establishes constraint-set nesting and an ordered global-minimum feature sequence under separate assumptions. It does not establish nested attraction basins, nor identify those sets with metastable phases. Relating this sequence to the dynamics studied in \citet{grokking} requires a separate calculation of local minima, barriers, and escape dynamics. The universality claim above is only partly tested: the transitions themselves are observed in nonlinear networks \citep{ersoy1,ersoy2}, but tracking the Hessian or Fisher spectrum through training of a nonlinear network, and checking which feature-aligned directions gain statistical distinguishability, is a natural experiment not carried out here. 

\paragraph{Outlook.}
The ordered feature subsets of Sec.~\ref{sec:nested} offer labels for comparing the metastable states studied in \citet{grokking}. Whether their attraction basins have a corresponding organization remains open. Feature acquisition, phase selection, and gain of Fisher distinguishability are related in the exact models under the stated conditions; their coincidence, and its relation to memorization and generalization, must be tested in other settings. Whether an approximate version of $G_d$ can be written down past the linear, aligned-Gaussian setting in general nonlinear networks, where the transitions are observed \citep{ersoy1,ersoy2} and the general symmetry-constraint machinery of \citet{ziyin-sym,zxpc} already applies, is the key question for turning the general claim into a quantitative theory of how networks acquire features. The two phenomena of Sec.~\ref{sec:detecting} are what make that question answerable empirically rather than only analytically.
\appendix

\section{The coefficient-preserving group}
\label{app:group}
Write $m=n_{\mathrm{out}}$ and $n=n_{\mathrm{in}}$. Preserving $u_i^{\top}Mv_i$ for every $M\in\mathbb R^{m\times n}$ is equivalent to
\begin{equation}
 (g^{\top}u_i)(h^{\top}v_i)^{\top}=u_iv_i^{\top}.
 \label{eq:app-coefficient-preservation}
\end{equation}
Since both sides have rank one, there is a nonzero scalar $\lambda_i$ such that $g^{\top}u_i=\lambda_i u_i$ and $h^{\top}v_i=\lambda_i^{-1}v_i$. In orthonormal bases beginning with the $d$ constrained directions, the group therefore consists of
\begin{equation}
 g=\begin{pmatrix}\Lambda&0\\ A&B\end{pmatrix},\qquad
 h=\begin{pmatrix}\Lambda^{-1}&0\\ C&D\end{pmatrix},
 \label{eq:app-group-blocks}
\end{equation}
where $\Lambda$ is invertible and diagonal, $B\in\GL(m-d)$, $D\in\GL(n-d)$, and $A,C$ are unrestricted. Counting these entries gives
\begin{equation}
 \dim G_d=d+m(m-d)+n(n-d).
 \label{eq:app-group-dimension}
\end{equation}
This counts transformations preserving coefficient functionals, not independent orbit tangents at a particular configuration. In particular, the decrease of group dimension is not the number of Fisher modes gained by imposing one scalar constraint. To describe the action used in Sec.~\ref{sec:sectors}, additionally take a canonical aligned rank-$d$ factorization, with nonzero product block $P=\operatorname{diag}(p_1,\ldots,p_d)$ and all other factor blocks zero. The infinitesimal action on its product has the form
\begin{equation}
 \delta M=\begin{pmatrix}0&PC_0^{\top}\\ A_0P&0\end{pmatrix},
 \label{eq:app-contingent-tangent}
\end{equation}
where $A_0,C_0$ are the infinitesimal lower-left blocks. Complement-only transformations fix the zero factor blocks; reciprocal rescalings of active pairs induce architectural gauge tangents; and the displayed mixing blocks change the represented map. The coefficient constraints alone do not require this canonical rank-$d$ configuration.

\paragraph{Norm-preserving restrictions.}
\label{app:mse}
Intersecting $G_d$ with $\OG(m)\times\OG(n)$ sets $A=C=0$, makes $B,D$ orthogonal, and restricts $\Lambda$ to diagonal signs. This is the restriction $O_d$ in Fig.~\ref{fig:groups}. For orthogonal $g,h$, invariance of the full map error $\tfrac12\lVert M-S\rVert_F^2$ for every $M$ additionally requires $gSh^{\top}=S$. Neither restriction is the architectural hidden-unit action $(W_2,W_1)\mapsto(W_2G,G^{-1}W_1)$, which preserves $M$ itself.

\section{Hessian sectors and Fisher information}
\label{app:goldstone}
\label{app:fisher}
Consider one hidden layer of width $w$, whitened inputs, and
\begin{equation}
 \mathcal L_\beta=	\frac{1}{2}\lVert W_2W_1-S\rVert_F^2
 +\tfrac\beta2\bigl(\lVert W_2\rVert_F^2+\lVert W_1\rVert_F^2\bigr),
 \qquad S=\Syx,\quad \beta>0.
 \label{eq:app-matrix-loss}
\end{equation}
The original population squared error differs from its first term only by a parameter-independent constant. Let $v=(U,V)$ be a perturbation of $(W_2,W_1)$, $R=W_2W_1-S$, and $J_Mv=UW_1+W_2V$. Expanding the product to second order gives
\begin{equation}
 v^{\top}H_{\mathcal L}v
 =\lVert J_Mv\rVert_F^2+2\langle R,UV\rangle_F
   +\beta\bigl(\lVert U\rVert_F^2+\lVert V\rVert_F^2\bigr).
 \label{eq:app-hessian}
\end{equation}
For $y\mid x\sim\mathcal N(Mx,\varsigma^2 I_m)$ with $\mathbb E[xx^{\top}]=I_n$, the expected Fisher quadratic form is
\begin{equation}
 v^{\top}\Fisher v=\varsigma^{-2}\lVert J_Mv\rVert_F^2.
 \label{eq:app-fisher}
\end{equation}
Thus neither the residual term nor the penalty term belongs to the Fisher matrix; at $R=0$, the unregularized error Hessian equals $\varsigma^2\Fisher$ \citep{martens}. Now evaluate at a canonical balanced rank-$d$ critical point, whose active factor blocks are $\operatorname{diag}(\sqrt{p_a})$ with $p_a=\eta_a-\beta>0$, and whose other blocks vanish. A gauge tangent $V_K=(W_2K,-KW_1)$ satisfies $J_MV_K=0$. Differentiating the gauge transformation of the error gradient and using $\nabla E=-\beta\theta$ at the critical point yields
\begin{equation}
 H_{\mathcal L}V_K=\beta V_{K+K^{\top}}.
 \label{eq:app-gauge-hessian}
\end{equation}
Skew-symmetric $K$ therefore produces eigenvalue zero, and symmetric $K$ produces eigenvalue $2\beta$, whenever the induced tangent is nonzero. Balancedness makes these two tangent spaces orthogonal. In either case, matrices supported only on the inactive hidden block induce zero tangents and must be excluded from the count.

\begin{center}
\small
\begin{tabular}{@{}llll@{}}
\toprule
Sector & Dimension & $H_{\mathcal L}$ eigenvalue & Fisher norm \\
\midrule
Orthogonal gauge & $wd-d(d+1)/2$ & $0$ & zero \\
Symmetric gauge & $wd-d(d-1)/2$ & $2\beta$ & zero \\
Inactive-unit complement & $(w-d)(m+n-2d)$ & $\beta\pm\eta_b$, $\beta$ & zero \\
Function-changing & $d(m+n-d)$ & data-dependent & positive \\
\bottomrule
\end{tabular}
\end{center}

The third sector couples inactive hidden units to the unrepresented input and output complements. Its first-order product differential is zero, and its quadratic blocks are Eq.~\eqref{eq:onset-hessian}. Writing $r=\rank S$, each omitted nonzero singular value supplies $\beta\pm\eta_b$, each with multiplicity $w-d$; the remaining $(w-d)(m+n-2r)$ directions have eigenvalue $\beta$. Here $d$ active modes are selected from the $r$ signal directions. The last sector is $(\ker J_M)^\perp$: the image of $J_M$ has arbitrary active-active and active-complement blocks, and a zero complement-complement block. Consequently
\begin{equation}
 \rank\Fisher=\rank J_M=d(m+n-d).
 \label{eq:app-fisher-rank}
\end{equation}
The four dimensions sum to $w(m+n)$. The table separates geometric subspaces, not necessarily distinct numerical eigenvalues; additional Hessian zero modes occur at thresholds or degeneracies. At saddles,
Fisher-positive directions need not have positive Hessian eigenvalues. The detailed spectrum is treated in \citet{cascade}. For the symmetric gauge sector, Eq.~\eqref{eq:app-gauge-hessian} also gives $H_EV_K=\beta V_K$: its total curvature $2\beta$ contains equal contributions from the error Hessian and the penalty Hessian.

\section{Ordered thresholds in the aligned deep-linear model}
\label{app:thresholds}
For $N\geq2$ scalar factors per aligned feature and a fixed positive product $p$, the arithmetic-geometric mean inequality gives $\sum_{\ell=1}^N a_\ell^2\geq Np^{2/N}$, with equality for equal factor magnitudes. Each feature therefore reduces to
\begin{equation}
 F_N(p)=\tfrac12(p-\eta)^2+\tfrac{\beta N}{2}p^{2/N},\qquad p\geq0.
 \label{eq:app-deep-reduction}
\end{equation}
For $N=2$, $p^*=(\eta-\beta)_+$ and the transition is continuous at $\beta_c=\eta$. For $N>2$, solving $F_N'(p_c)=0$ together with $F_N(p_c)=F_N(0)$ gives
\begin{equation}
 p_c=\frac{N-2}{N-1}\eta,\qquad
 \beta_c=\frac{\eta}{N-1}
 \left(\frac{N-2}{N-1}\eta\right)^{(N-2)/N}.
 \label{eq:app-deep-threshold}
\end{equation}
For $N>2$ and $p>0$, the sign of $F_N(p)-F_N(0)$ is the sign of $\beta-(2\eta p-p^2)/(Np^{2/N})$. The latter ratio has its unique positive maximum at $p_c$ on $0<p<2\eta$, so the displayed value is the global activation threshold. The nonzero jump is first order, and $\beta_c$ increases strictly with $\eta$. Thus distinct strengths and a uniform penalty order the global activation thresholds in the separable aligned model of Sec.~\ref{sec:nested}. This comparison does not determine which local minimum an optimizer reaches. More generally, for any nonnegative error finite at zero, a global minimizer of $E+\beta\lVert\theta\rVert^2/2$ obeys $\lVert\theta_\beta\rVert^2\leq2E(0)/\beta$, establishing convergence to zero as $\beta\to\infty$ when all parameters are penalized.

\section{Numerical construction of the figures}
\label{app:numerics}

\paragraph{Exact landscapes.}
Figures~\ref{fig:filtration} and~\ref{fig:cascade} evaluate the displayed bilinear losses directly on uniform factor-space grids in $[-2,2]^2$, with $\eta=(1.2,0.8)$ and respectively 401 and 400 points per axis. The coefficient-space filtration panels use 401 points per axis on $[-0.55,2.15]^2$. The filtration uses $(\rho_1,\rho_2)=(1,0),(1,1)$; an omitted feature is represented by $(a_i,b_i)=(0.6,0.2)$, while an imposed feature uses $a_i=b_i=\sqrt{\eta_i}$. The regularized panels use strengths $(0.3,1.2,2.0)$ for feature 1 and $(0.2,0.8,1.5)$ for feature 2. They mark the positive balanced minimum $a_i=b_i=\sqrt{(\eta_i-\beta)_+}$; the plotting code denotes $\beta$ by $\lambda$. No training trajectories are used for these figures.

\paragraph{Illustrative rough landscapes.}
For Fig.~\ref{fig:bbp}, the plotting code evaluates
\begin{equation}
 E_T(p)=\tfrac12\sum_{i=1}^2\rho_i(T)(p_i-\eta_i)^2+A(T)\xi(p),\qquad
 A(T)=0.055+1.15(T/1.2)^{4.5},
 \label{eq:app-rough-loss}
\end{equation}
where $\rho_i(T)=(1-T^2/\eta_i^2)_+$ and $\xi$ is one fixed random Fourier field. Its raw form is $\sum_{j=1}^{85}a_j\cos(k_j\cdot p+\phi_j) +\sum_{j=1}^{70}\widetilde a_j\cos(\widetilde k_j\cdot p+ \widetilde\phi_j)$. Directions are sampled by normalizing independent two-dimensional standard Gaussian vectors; frequency magnitudes are log-uniform on $[0.8,9]$ and $[7,20]$, respectively. Phases are uniform on $[0,2\pi)$, and amplitudes are independent Gaussians with standard deviations $85^{-1/2}$ and $0.35\,70^{-1/2}$. The code uses NumPy's \texttt{default\_rng(14)} and normalizes this field once to zero mean and unit standard deviation on a $220\times220$ grid over $[-10.25,10.35]\times[-10.25,10]$. At $T=1.68,1.02,0.60$, $p^*$ is the minimum on a $360\times360$ grid over the same window, without continuous refinement. The factor panels evaluate $E_T(a_1b_1,p_2^*)$ or $E_T(p_1^*,a_2b_2)$ on $360\times360$ grids in $[-5,5]^2$. Logarithmic shading uses each panel's excess loss $E_T-\min E_T+\epsilon$, with $\epsilon=\max(10^{-8},10^{-3}q_{0.9})$ and upper cutoff $q_{0.9}+\epsilon$, where $q_{0.9}$ is the 90th percentile of that excess loss. This additional random field is a visualization model, not a sampled cross-covariance or a consequence of BBP theory. It is absent from the phase-diagram calculation below.

\paragraph{Analytic phase diagram.}
\label{app:phase}
Figure~\ref{fig:phase} uses Eq.~\eqref{eq:phase-loss}, with $\eta=(1.2,0.8)$ and $s=(3,1)$, on logarithmically spaced grids of
100 values each: $T\in[0.05,1.60]$ and $\beta\in[0.005,2]$. Since $a_i^2+b_i^2\geq2|a_ib_i|$, minimization at fixed product reduces
each feature to $\tfrac12\rho_i(p_i-\eta_i)^2+\beta s_i|p_i|$. Panel (a) classifies its global minimum using $\beta<\beta_c^{(i)}(T)=\rho_i(T)\eta_i/s_i$. The analytic crossing is
\begin{equation}
 T_\times=\sqrt{72/175}\simeq0.6414,\qquad \beta_\times=2/7\simeq0.2857.
 \label{eq:app-phase-crossing}
\end{equation}

\paragraph{Finite-time numerical panel.}
Panel (b) uses one finite-time trajectory per grid point. At each step, the parameters undergo a gradient-descent update followed by an independent Gaussian kick, as specified in
Eq.~\eqref{eq:langevin-protocol}. This is the Euler-Maruyama discretization of overdamped Langevin dynamics \citep{higham2001}. We use a fixed noise strength $T_L=10^{-2}$, a time step $\Delta t=10^{-3}$, and 12,000 updates. Initial parameter components are drawn independently from $\mathcal N(0,0.1^2)$. The parameter-noise strength $T_L$ is held constant throughout the diagram; the plotted variable $T$ controls the prescribed feature weights $\rho_i(T)$. The gradients used are
\begin{equation}
 \partial_{a_i}\mathcal L_{\beta,T} =\rho_i(p_i-\eta_i)b_i+\beta s_i a_i,\qquad \partial_{b_i}\mathcal L_{\beta,T} =\rho_i(p_i-\eta_i)a_i+\beta s_i b_i.
 \label{eq:app-phase-gradients}
\end{equation}
All components are evaluated before updating the parameters. Products are recorded after updates whose zero-based loop indices
satisfy $n>4800$ and $n\equiv0\pmod{20}$, giving 359 samples per trajectory. Feature $i$ is active when its sample mean exceeds
$0.2\eta_i$. The implementation uses \texttt{default\_rng(0)}, traverses $T$ in the outer loop and $\beta$ in the inner loop, and assigns each cell its initialization followed by its Gaussian kicks. The dashed lines in both panels are the analytic activation boundaries. For the numerical panel's threshold applied to a deterministic minimum, the reference boundaries would instead be $0.8\beta_c^{(i)}(T)$, crossing at the same $T_\times$ and $\beta=8/35$. Consequently, criterion choice, finite time, parameter noise, and discretization can all contribute to differences between the panels. $T_L$ is fixed, no noisy datasets are sampled, and one trajectory per cell does not establish equilibration or sampling uncertainty.


\end{document}